\documentclass{aa}  

\usepackage{multirow}
\usepackage{graphicx}
\usepackage{color}

\usepackage{txfonts}
\usepackage{url}

\begin{document}

   \title{Star formation rates and stellar masses from machine learning}

   \author{V. Bonjean\inst{1,2}
          \and N. Aghanim\inst{1}
          \and P. Salomé\inst{2}
          \and A. Beelen\inst{1}
          \and M. Douspis\inst{1}
          \and E. Soubrié\inst{1}
          }

   \institute{Institut d'Astrophysique Spatiale (IAS), CNRS, Université Paris-Sud, UMR 8617 Bâtiment 121, Orsay, France\\
   \email{victor.bonjean@ias.u-psud.fr}
         \and
             LERMA, Observatoire de Paris, PSL Research University, CNRS, Sorbonne Universités, UPMC Univ. Paris 06, 75014, Paris, France\\
             }

   \date{Received XXX; accepted XXX}

  \abstract
  {Star-formation activity is a  key property to probe the structure  formation and hence characterise the large-scale structures of the universe. This information can be deduced from the star formation rate (SFR) and the stellar mass ($\mathrm{M}_\star$), both of which, but especially the SFR, are very complex to estimate. Determining these quantities from UV, optical, or IR luminosities relies on complex modeling and on priors on galaxy types. We propose a method based on the machine-learning algorithm Random Forest to estimate the SFR and the $\mathrm{M}_\star$ of galaxies at redshifts in the range $0.01<z<0.3$, independent of their type. The machine-learning algorithm takes as inputs the redshift, WISE luminosities, and WISE colours in near-IR, and is trained on spectra-extracted SFR and $\mathrm{M}_\star$ from the SDSS MPA-JHU DR8 catalogue as outputs. We show that our algorithm can accurately estimate SFR and $\mathrm{M}_\star$ with scatters of $\sigma_{\mathrm{SFR}}=0.38$ dex and $\sigma_{\mathrm{M}_\star}=0.16$ dex for SFR and stellar mass, respectively, and  that it is unbiased with respect to redshift or galaxy type. The full-sky coverage of the WISE satellite allows us to characterise the star-formation activity of all galaxies outside the Galactic mask with spectroscopic redshifts in the range $0.01<z<0.3$. The method can also be applied to photometric-redshift catalogues, with best scatters of $\sigma_{\mathrm{SFR}}=0.42$ dex and $\sigma_{\mathrm{M}_\star}=0.24$ dex obtained in the redshift range $0.1<z<0.3$.}


   \maketitle

\section{Introduction}\label{intro}

The galaxy types and their relations to the environment are key features of the study and characterisation of large-scale structures (LSS) in the context of future large surveys of galaxies such as the Large Synoptic Survey Telescope (LSST)\footnote{\url{https://www.lsst.org}}, the Dark Energy Survey (DES)\footnote{\url{https://www.darkenergysurvey.org}} or Euclid\footnote{\url{https://www.euclid-ec.org}}.

\begin{figure}[!ht]
    \centering
    \includegraphics[width=0.5\textwidth]{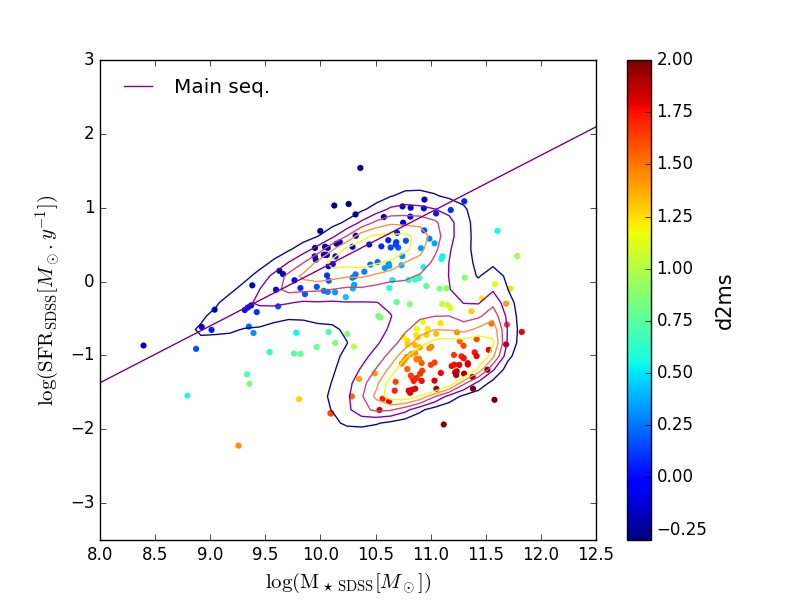}
    \caption{SFR vs. $\mathrm{M}_\star$ diagram. The contours represent the $1\sigma$ to $5\sigma$ isodensities of all the SDSS MPA-JHU DR8 values from the training sample. The dots are 100 random galaxies taken in the catalogue. The purple solid line is the main sequence of star forming galaxies given by \citep{elbaz}. The colours of the galaxies are a function of the distance to the main sequence, d2ms, and are directly representative of the passivity of the galaxies.}
    \label{d2ms}
\end{figure}

In the standard understanding of galaxy evolution, star-forming galaxies (usually blue and spiral ones) align along a main sequence in diagrams showing their star formation rates (SFR) versus their stellar mass ($\mathrm{M}_\star$) (blue dots in Fig.~\ref{d2ms}). This sequence has been fitted for low-redshift galaxies (up to $z\sim0.3$) by \cite{brinchmann} using the \textit{Sloan Digital Sky Survey} \citep[SDSS,][]{sdss} galaxies \citep{elbaz} : $\mathrm{SFR}_{\mathrm{SDSS}}\left[\mathrm{M}_\odot.y^{-1}\right]=8.7\times{\left[ \mathrm{M}_\star/10^{11} \mathrm{M}_\odot\right]}^{0.77}.$

Galaxies leave the main sequence when they stop forming stars (quenching), that is, when they loose their cold gas. This can be due to different processes that are not yet fully understood: harassment \citep[e.g.][]{moore1995}, strangulation \citep[e.g.][]{peng2015}, that is, when they enter a region with denser and hotter gas (e.g. galaxy clusters or inner parts of cosmic filaments), or ejection of the gas through AGN jets \citep[e.g.][]{dubois2013}. In all cases, galaxies stop forming stars and undergo a transitioning stage, the so-called green-valley \citep{alatalo2014}, and finally become passive (or red and dead) galaxies (red dots in Fig.~\ref{d2ms}). In this general picture, the activity of a galaxy is usually defined in terms of its SFR or of its stellar mass.

Estimating the quantities SFR and $\mathrm{M}_\star$ is complex (see \cite{kennicutt2012} for a review); they are directly or indirectly related to the observations of stars. We briefly review here the dependence of the star properties on wavelength across the electromagnetic spectrum. The young and massive O- and B-type stars are the hottest and thus the most energetic stars. Their blackbody spectra peak in the blue wavelength and they strongly emit in the UV. The UV luminosity of distant galaxies traces these types of stars that in turn directly relate to the SFR, as they represent the youngest stellar populations. However, at these wavelengths the dust absorption is very important and  correcting the UV luminosities from the dust attenuation is not trivial \citep{lagache2005, kennicutt2012}. Multi-wavelength tracers or dust attenuation estimations in the UV/optical \citep{calzetti1994, kennicutt1998, salim2007, kennicutt2012, janowiecki} are therefore needed to correct UV luminosities and use them as a direct tracer to derive estimations of SFR.

The non-ionizing low-mass old stars represent most of the contribution to the galaxy luminosities in the optical. As they are the most numerous in a galaxy, the optical luminosity is also directly related to the number of stars, and thus to the stellar mass, provided that there exists a theoretical model of star population and an initial mass function (IMF) \citep[e.g.][]{bruzual}. The estimation of stellar masses strongly depends on the assumed IMF. For example, a typical correcting/calibration factor of $\sim1.6$ is needed to change from a stellar mass with a Salpeter IMF \citep{salpeter} to a stellar mass with a Chabrier \citep{chabrier} IMF \citep{haas}.

In the near-IR (NIR) ($\sim0.8\mu$m$<\lambda<\sim3\mu$m), the old and non-massive stars also represent most of the contribution to the total luminosity. These wavelengths can therefore also trace the stellar mass through the old population, in the same way as optical measurements do \citep{wen2013}. In the mid-IR (MIR) ($\sim3\mu$m$<\lambda<\sim70\mu$m), the contribution of dust becomes predominant. Particularly in the 8-12 $\mu$m band, the contribution of heated small grains and polycyclic aromatic hydrocarbon (PAH, \cite{pah}) offers a useful tool to study the composition and the abundance of dust. From $\sim20\mu$m to $\sim70\mu$m, the luminosity is mostly due to thermalised dust and large grains heated by the UV emission of the energetic young O- and B-types stars. The luminosity in the IR is thus indirectly related to the SFR and this was performed using, for example, the $8\mu$m and the $24\mu$m bands from the Spitzer satellite \citep{spitzer} or the $12\mu$m and the $22\mu$m from the Wide-Field Infrared Survey Explorer \citep[WISE, ][]{wise} satellite \citep{calzetti, kennicutt2009, jarrett2013, cluver2014, cluver2017}.

All of these relations are well calibrated, but applying them to galaxies without having any prior on their types can lead to potential biases, as passive galaxies do not have the same properties in the IR (see Sect.~\ref{discussion}). Ideally, optical spectroscopic data are needed to estimate the SFR and $\mathrm{M}_\star$ properties, but they are not always available and are costly in terms of observing time. This is even more prohibitive when the goal is to characterise the galaxy properties in large surveys. 

In this study, we propose an alternative approach to estimate SFR and $\mathrm{M}_\star$ for all galaxies over 70\% of the sky (i.e. outside the Galactic plane) with measured redshifts in the range $0<z<0.3$ (the redshift limit of the training catalogue), without any priors on galaxy types. To do so, we use a machine-learning algorithm. As a matter of fact and for several years already, machine-learning algorithms have been developed and have now become reliable tools to classify or estimate physical properties of astrophysical objects \citep{aghanim2015, marc2015, bilicki2014, bilicki2016, krakowski, rfsimu, pashchenko, domingez, tuccillo, ucci, viqar2018, veneri2018, vipers2018a, vipers2018b}. The development of the \texttt{scikit-learn} library \citep{sklearn} in \texttt{Python} has made them relatively easy to use. It has allowed `pythoneer' astrophysicists to develop and test their own machine-learning algorithms on their data. 

In the machine learning domain, algorithms are designed either to estimate or classify features based on reference samples, or to identify commonalities on the input features without resorting to any models. These two families of algorithms of machine learning are called supervised and unsupervised algorithms, respectively. The first include algorithms such as Multi-Layer Perceptron, Random Forests, Support Vector Machine, Deep learning, and so on. The second include clustering methods such as k-mean algorithms (see the \texttt{scikit-learn} website\footnote{\url{http://scikit-learn.org/}} for more details about the algorithms). In our analysis, we use a supervised machine-learning algorithm. Such a method is able to estimate very non-linear laws based on models trained on reliable given inputs and outputs. In the present case, it allows us to estimate SFR and $\mathrm{M}_\star$ independently of any complex model or any priors on galaxy types. The quality and pertinence of the results of a supervised machine-learning algorithm depend greatly on the training set and how it captures the features that will be classified. 

In Sect.~\ref{data}, we therefore present the data used to generate the trained model and train the machine learning algorithm. In Sect.~\ref{sect3}, we present an analysis of the Random Forest algorithm, and the results. We discuss the results and the limitations of the method in Sect.~\ref{discussion}, and also present an example and an illustration of the method. Finally, we briefly summarise the method in Sect.~\ref{summary}.

We use the {\it Planck} 2015 cosmological parameters throughout this paper \citep{planckcosmo} with $\mathrm{H}_0=67.74$ $\mathrm{km.Mpc^{-1}.s^{-1}}$, ${\Omega_\mathrm{M}}_0=0.3075$ and ${\Omega_\mathrm{b}}_0=0.0486$. Also, all luminosities noted as $\mathrm{L}\nu$ in different wavelengths refer to the luminosity density in this wavelength $\nu L_{\nu}$.

\begin{figure}[!ht]
    \centering
    \includegraphics[width=0.5\textwidth]{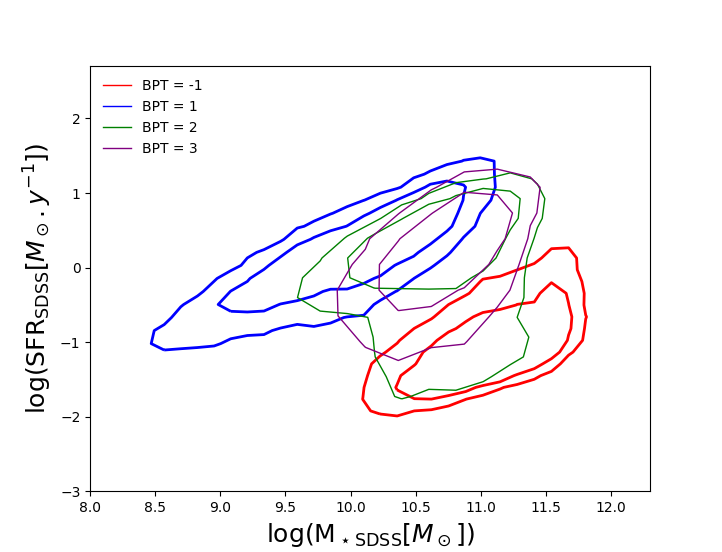}
    \caption{SFR and $\mathrm{M}_\star$ provided in the SDSS catalogue. Each contour represents a galaxy population as defined by their position on the BPT diagram. Red contours represent passive galaxies, blue contours star-forming galaxies, green contours the galaxies from the green valley, and purple contours the AGNs.}
    \label{probe_bpt}
\end{figure}

\section{Data}\label{data}

In this section, we present the data used to construct the training catalogue of the machine-learning algorithm.

\subsection{WISE}

The WISE satellite surveyed the whole sky in four near- and mid-infrared wavelengths (3.4, 4.6, 12 and 22$\mu$m). From the coadded WISE Atlas Images (at angular resolutions respectively of 6.1, 6.4, 6.5 and 12 arcsec), the AllWISE Source Catalogue\footnote{Available at \url{http://wise2.ipac.caltech.edu/docs/release/allwise/expsup/sec1_3.html#src_cat}} was generated with accurate positions, photometry, and ancillary information for 747,634,026 detected sources \citep{cutri2013}. For our study, we use the profile-fitted photometry measurements of the W1 (3.4$\mu$m), W2 (4.6$\mu$m), and W3 (12$\mu$m) bands, noted \textsf{w1mpro}, \textsf{w2mpro,} and \textsf{w3mpro} in the AllWISE Source Catalogue. The associated errors and signal-to-noise ratios (S/Ns) of magnitude measurements are noted \textsf{w1sigmpro}, \textsf{w2sigmpro}, \textsf{w3sigmpro}, and \textsf{w1snr}, \textsf{w2snr}, \textsf{w3snr,} respectively. We reject the sources with known detection or measurement artifacts by taking \textsf{cc\_flags} = 0 for each of the three bands. We also select the sources with high-quality photometry measurements. The WISE magnitudes being upper-values below \textsf{w*snr} < 2 \citep{krakowski}, where \textsf{*} can be \textsf{1}, \textsf{2,} or \textsf{3}, we proceed in the same way as \cite{krakowski} and select only sources with reliable magnitudes in W1 and W2: \textsf{w1snr} > 2, \textsf{w2snr} > 2. Around one third of the selected sources have \textsf{w3snr} < 2. In this case, the \textsf{w3mpro} upper-value magnitudes can be considered as a typical bias, and a correction of +0.75 can be applied to the values to correct for this bias (Fig.~A.1 in \cite{krakowski}). We also apply a $0^{\mathrm{th}}$-order k-correction (dependence on redshift) by adding the quantity $-2.5\times \mathrm{log}\left(1+z\right)$ to the measured magnitudes in each of the three bands, where we take the spectroscopic redshift $z$ from the SDSS catalogue (see the cross-match procedure between the two catalogues in Sect.~\ref{xmatch}).

\subsection{SDSS}

The SDSS is one of the largest optical surveys available. It has produced deep images of one third of the sky in five optical bands: u, g, r, i, and z, and has performed spectroscopic measurements for more than three million astronomical objects. From these data, several value-added catalogues were generated, which provide a wealth of information about the objects thanks to the study of a large panel of spectral emission lines. We use here the MPA–JHU DR8 catalogue, from the Max Planck Institute for Astrophysics and the Johns Hopkins University \citep{kauffmann, brinchmann}. It provides SFR and stellar masses for 1,843,200 galaxies with redshifts up to $z\sim0.3$ (Fig.~\ref{probe_bpt}). These data based on the SDSS DR8 release are publicly available\footnote{\url{http://sdss3.org/dr8/}} together with all details about the catalogue and the computations and fits of the galaxy physical properties. 

The SFR (flagged as \textsf{SFR\_TOT\_P50}) are estimated using the $\mathrm{H}_\alpha$ emission lines (when available) corrected from the dust extinction with the Balmer decrement $\mathrm{H}_\alpha/\mathrm{H}_\beta$ \citep{brinchmann}. For no-emission line galaxies, SFRs are estimated using a relation between the SFR and the spectral index $\mathrm{D}_{4000}$ \citep{bruzual1983, balogh1999, brinchmann}. The $\mathrm{M}_\star$ (flagged as \textsf{LGM\_TOT\_P50}) are computed based on theoretical models of stellar populations \citep{kauffmann}, and assuming a Kroupa IMF \citep{kroupa}.

The SDSS MPA-JHU DR8 catalogue provides BPT classes (flagged as \textsf{BPTCLASS}), which depend on the position of the galaxies in the Baldwin, Phillips, \& Terlevich (BPT) diagram \citep{bpt}. This diagram can segregate a population of galaxies by comparing the emission-line ratios $\left[\mathrm{OIII}\right]/\mathrm{H}_\beta$ and $\left[\mathrm{NII}\right]/\mathrm{H}_\alpha$ (see Fig.~\ref{probe_bpt}). In the classification provided by the MPA-JHU catalogue, \textsf{BPTCLASS} = 1 corresponds to star-forming galaxies, \textsf{BPTCLASS} = 2 to composite galaxies (transitioning), \textsf{BPTCLASS} = 3 to AGNs, and \textsf{BPTCLASS} = 4 and \textsf{BPTCLASS} = 5 to low-S/N emission line galaxies \citep{brinchmann}. The class \textsf{BPTCLASS} = -1 corresponds to galaxies unclassifiable in the BPT diagram: passive galaxies without emission lines \citep{brinchmann}.

From the SDSS catalogue, we generate a purer catalogue by selecting objects with only reliable properties. To do so we set the following flags: \textsf{RELIABLE} $\ne 0$, \textsf{Z\_WARNING} $= 0$, \textsf{SFR\_TOT\_P50} $\ne -9999$, \textsf{LGM\_TOT\_P50} $\ne -9999$, and \textsf{Z} $> 0$. This pure catalogue contains 794 633 galaxies.

\begin{figure}[!ht]
    \centering
    \includegraphics[width=0.5\textwidth]{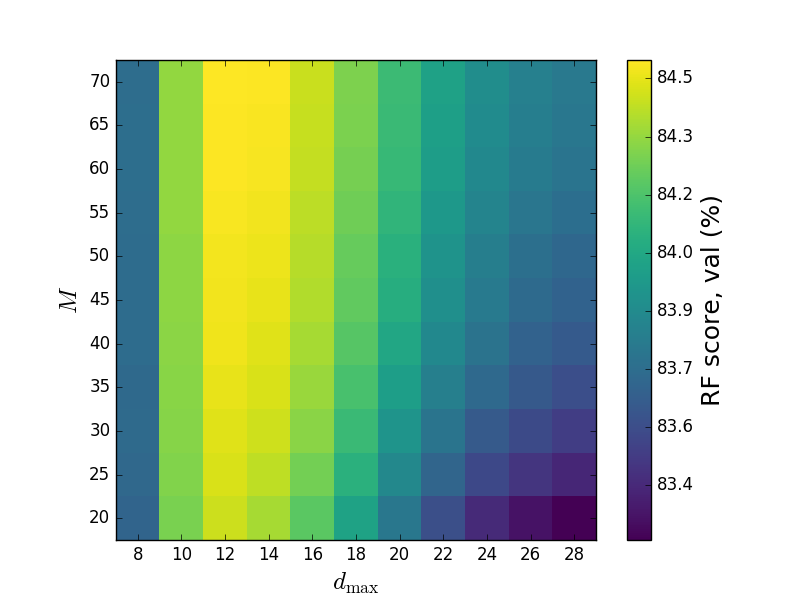}
    \caption{Percentage score of the RF results on the validation sample as a function of the RF parameters $M$ and $d_\mathrm{max}$ ($M$ being the number of trees and $d_\mathrm{max}$ the maximum depth). Setting $M=40$ and $d_\mathrm{max}=12$ is enough in our case to optimise the RF.}
    \label{2d_score}
\end{figure}

\begin{figure*}[!ht]
    \centering
    \includegraphics[width=0.5\textwidth]{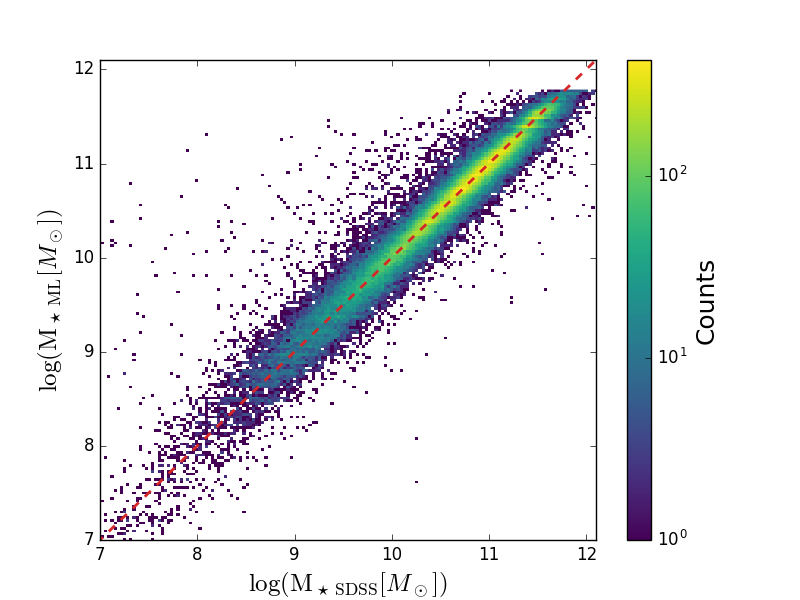}\includegraphics[width=0.5\textwidth]{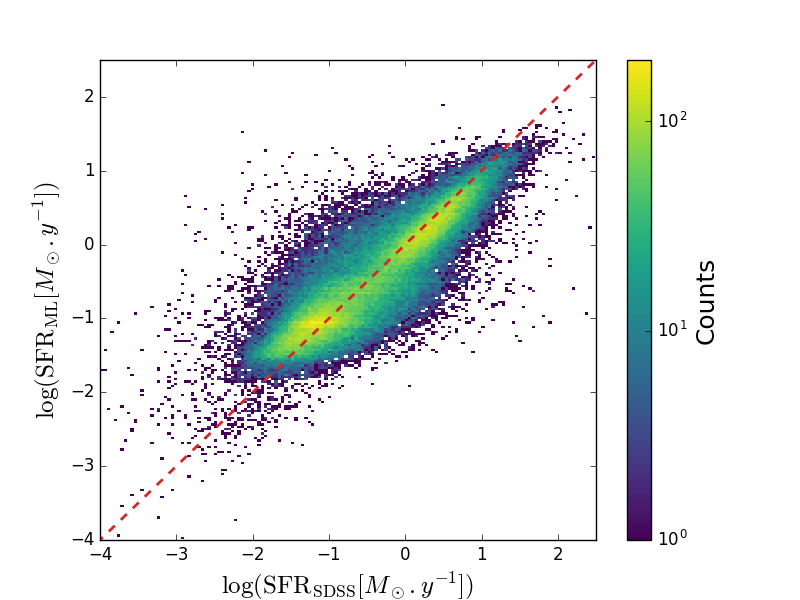}
    \caption{Results of the RF on the test sample (20\% of the entire sample), with optimisation parameters set to $M=40$ and $d_\mathrm{max}=12$. Left: $\mathrm{M}_\star$ estimate with the RF compared to $\mathrm{M}_\star$ from the SDSS MPA-JHU DR8 catalogue. Right: SFR estimate with the RF compared to SFR from the SDSS MPA-JHU DR8 catalogue.}
    \label{results}
\end{figure*}

\section{The machine-learning algorithm}\label{sect3}

\subsection{Principle and advantages}\label{sect41}

Increasingly used in astrophysics and cosmology, machine-learning algorithms have become very powerful tools to detect, classify, or characterise astrophysical sources \citep[e.g. as a very non-exhaustive list:][and references therein]{aghanim2015, bilicki2014, bilicki2016, krakowski}. One of the main advantages of this technique is that  a model is not needed (usually complex or empirical) to perform regression on a set of data. Here only a set of input data and output data are needed; and the machine learns the relation (which can be very non-linear and complex) between the input and output data. Different kinds of machine-learning algorithms have been developed and are easily usable (e.g. see the \texttt{scikit-learn} website\footnote{\url{http://scikit-learn.org/}}). For this study, we choose to use the Random Forest (RF) algorithm \citep{ho1998} which is among the simplest, fastest to run, and easiest to understand among the many machine-learning methods (see Sect.~\ref{sect42}), and \texttt{scikit-learn} v.0.19.1. 

The usual way to estimate the efficiency of the algorithm (its ability to perform a regression) when applied to a training sample (i.e. the inputs and outputs data set) is to split this training sample into several subsamples, and train and test the algorithm on these different samples. One can then train the machine-learning algorithm on a subsample (50\% of the whole sample) and apply it to the other subsample (the other 50\%). Knowing the reference values of the last subsample and having their estimation by the machine-learning algorithm, the errors and the bias can be estimated by comparing the two.

\subsection{The choice of inputs and outputs}

Ensuring good, that is, unbiased, training of the algorithm, the choice of reference inputs and the outputs is essential.  This can be overridden by other algorithms, in particular deep-learning algorithms. In this study, the choice of using the WISE data is motivated by its full-sky coverage and its very large number of sources.

First of all, we have to define our inputs, that is, the data that will be used to estimate the SFRs and the stellar masses at the end. As the SFR can evolve with redshift, we choose to use $z$ as an input. As a proxy for the stellar mass, we use the luminosity in the W1 band ($3.4\mu$m) of WISE that traces the old non-ionizing stars \citep{wen2013, jarrett2013}. As a proxy for the SFR, we use the luminosity in the W3 band ($12\mu$m) of WISE, which traces the emission from small grains and is directly related to the total quantity of dust \citep{jarrett2013, cluver2014, cluver2017}. Although the W4 band of WISE is also a good tracer of the SFR \citep{jarrett2013, cluver2014, cluver2017}, its larger beam size of 12'' and its poorer sensitivity could lead to an important incompleteness and a significant bias of source selection with respect to redshift. We therefore decided not to use it. As we want to estimate the SFR and the stellar mass for both galaxy types (active and passive) without any prior, we also chose as input the two colours of WISE that can segregate the galaxy types: W1-W2 ($3.4-4.6\mu$m) and W2-W3 ($4.6-12\mu$m) \citep{wise}. 

We then needed to choose what outputs would be used as reliable reference for SFR and $\mathrm{M}_\star$. We chose to use the SFRs and stellar masses from the SDSS MPA-JHU DR8 catalogue, since their values are based on calibrated spectra \citep{brinchmann}. These latter authors estimated SFR for different types of galaxies using the metal lines as tracers (the $\mathrm{H}_\alpha$ recombination line for most of galaxies, corrected from the dust attenuation with the Balmer decrement $\mathrm{H}_\alpha/\mathrm{H}_\beta$), and the relation with the spectral index $\mathrm{D}_{4000}$ \citep{bruzual1983, balogh1999} for no-emission-lines galaxies. To estimate the stellar mass, \cite{brinchmann} used theoretical models of star populations fitted with Monte Carlo \citep{kauffmann} based on models from \cite{bruzual}, and a Kroupa IMF \citep{kroupa}.

\begin{figure*}[!ht]
    \centering
    \includegraphics[width=0.5\textwidth]{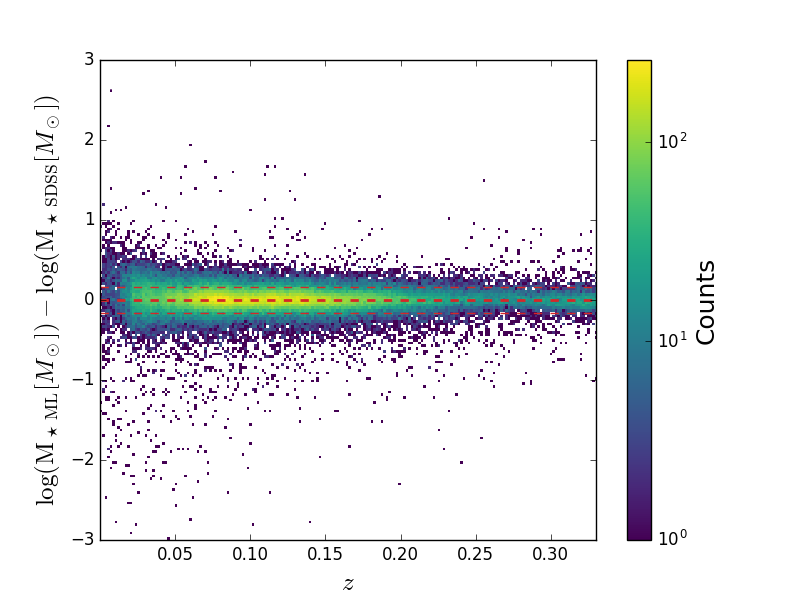}\includegraphics[width=0.5\textwidth]{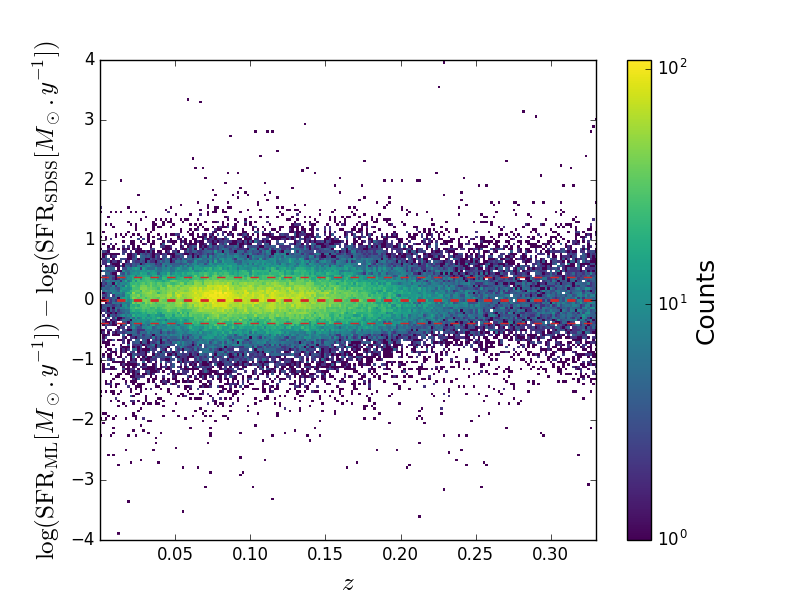}
    \caption{Errors of the RF results obtained for the test sample (same errors as those presented in Fig.~\ref{results}) as a function of redshift for $\mathrm{M}_\star$ and SFR. }
    \label{along_red}
\end{figure*}

\subsection{Constructing the training catalogue}\label{xmatch}

\begin{figure*}[!ht]
    \centering
    \includegraphics[width=0.5\textwidth]{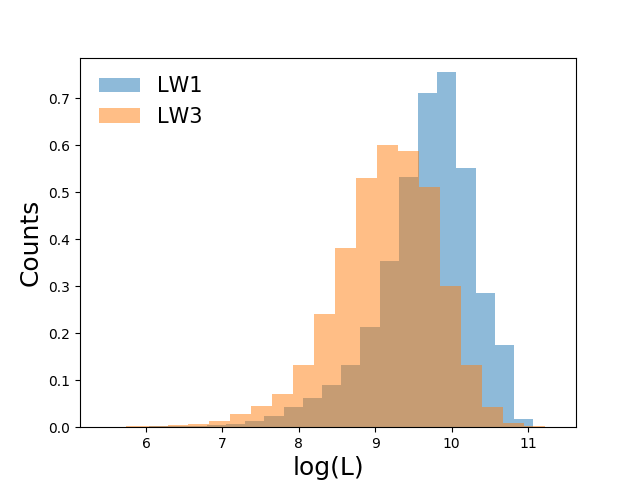}\includegraphics[width=0.5\textwidth]{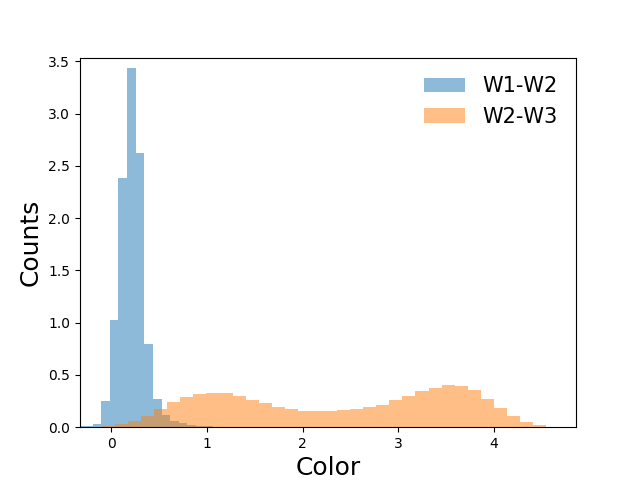}
    \caption{Histograms showing the range of the input data (luminosities and colours) for the sources of the training catalogue.}
    \label{histo_to}
\end{figure*}

We construct the training set by performing a positional cross-match of the SDSS subsample of 794,633 galaxies described in Sect.~\ref{data} with the AllWISE Sources Catalogue within a radius of 6'' (the beam of the W1 band of WISE from which the source positions are extracted). In order to ensure a pure catalogue, we remove all cross-match cases with multiple associations and end up with 603,293 galaxies. After removing sources with bad WISE magnitude measurements (only $\sim5\%$ of the total sources) as explained in Sect.~\ref{data}, we finally end up with a reliable catalogue of 573,582 galaxies. For all these galaxies, we have access to measured or estimated redshifts, WISE \textsf{w*mpro} magnitudes, SFRs, and $\mathrm{M}_\star$. This is the basis of our input and output data (see previous section). We show in Fig.~\ref{histo_to} the range of LW1, LW3, W1-W2 and W2-W3 (i.e. the inputs data) on the training catalogue.

\subsection{The Random Forest algorithm}\label{sect42}

The  RF algorithm used in the present study is based on decision tree learning. It uses a decision tree that splits the training set optimally by reducing the Gini impurity\footnote{detailed here:  \url{https://scikit-learn.org/stable/modules/tree.html#classification-criteria}}. The principle is to define if-else rules on the input features, in order to finally obtain the best and purest representation of the sets at each splitting, according to the outputs (see the documentation in the \texttt{scikit-learn} website\footnote{\url{http://scikit-learn.org/}}).

The RF algorithm uses the mean estimator of a sample of decision trees learned by bootstrapping the training set. For a training set of $n$ samples, with $X=x_1,...,x_n$ and $Y=y_1,...,y_n$, the inputs and the outputs of the machine learning, respectively, the estimator for an untrained value $x'$ is computed as

\begin{equation}\label{eq1}
    \tilde{y}\left(x'\right)=\frac{1}{M}\sum_{m=1}^M \tilde{y}_m\left(x'\right),
\end{equation}

where $M$ is the number of decision trees, and $\tilde{y}_m$ is the estimator for $x'$ of the decision tree $m$ trained on a random sample with replacement of $n$ elements in the sample of couples $\left(X, Y\right)$.

To optimise the results, some of the parameters have to be fixed, such as the number of trees $M$ or the maximum depth of the tree (i.e. the maximum number of splitting) $d_\mathrm{max}$. These parameters can be set by splitting the training set into several subsample sets, and by training and computing the score of the algorithm on these independent subsample sets. The best optimised parameters can be set when the score on independent subsample tests is the highest.

\subsection{Optimisation}

To set the optimal number of trees $M$ and the maximum depth $d_\mathrm{max}$, and further estimate the errors on the RF, we proceed in splitting the training set into subsamples (see Sect.~\ref{sect41}). Following standard procedures, we split our sample into three randomised subsamples: 60\% as the training set, 20\% as the validation set, and 20\% as the test set (we further check that changing the sizes of the subsamples does not affect the results of the algorithm). We train the RF on the training set, varying $M$ and $d_\mathrm{max}$. We compute the score of the RF on the validation set using the coefficient of determination $100\times\mathrm{R}^2$, where $\mathrm{R}^2=1-\sigma_\mathrm{res}^2/\sigma^2$, $\sigma_\mathrm{res}^2=\sum_{i=0}^n\left(\tilde{y}(x_i)-y_i\right)^2$ , which is the residual sum of squares, and $\sigma^2=\sum_{i=0}^n\left(y_i-\bar{y}\right)^2$ , the variance of the output distribution. We then set the two optimised parameters to the ones that give the highest scores on the validation set. Figure~\ref{2d_score} shows the score of the RF on the validation set, depending on the two parameters $M$ and $d_\mathrm{max}$. For the parameter $d_\mathrm{max}$, the performance of the RF increases until $d_\mathrm{max}=12$ and starts decreasing beyond. For the parameter $M$, a simple lower limit is enough to optimise the parameter, as increasing the value of $M$ to higher values does not give better results. Here, setting $M=40$ and $d_\mathrm{max}=12$ is sufficient to optimise the algorithm, for an optimal score of 84.5\% on the validation set.

\begin{figure*}[!ht]
    \centering
    \includegraphics[width=0.5\textwidth]{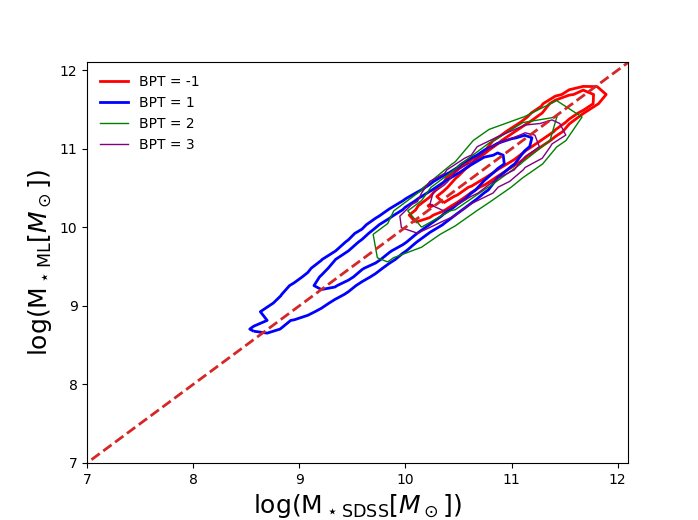}\includegraphics[width=0.5\textwidth]{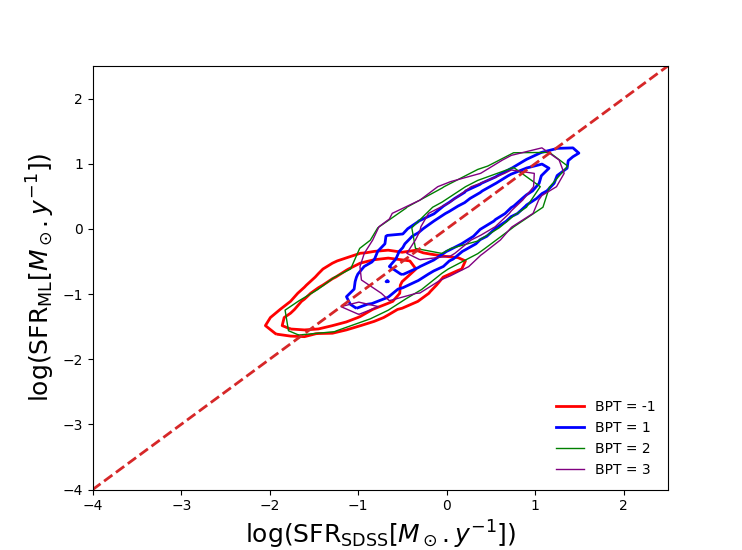}
    \caption{SFR and $\mathrm{M}_\star$ obtained with the RF algorithm on the test set compared to the SDSS classification based on the BPT diagram. Colour code of the contours is the same as in Fig.~\ref{probe_bpt}.}
    \label{BPT}
\end{figure*}


\subsection{Results and errors}\label{sec:test} 

We train the RF on the training set, with the two parameters fixed at $M=40$ and $d_\mathrm{max}=12$, and we estimate SFR and $\mathrm{M}_\star$ with the RF on the test set ($\mathrm{SFR}_\mathrm{ML}$ and ${\mathrm{M}_\star}_\mathrm{ML}$). We then compare these results with their reference values, defined as those of the SDSS catalogue. We can hence estimate the performance of the machine-learning algorithm in terms of errors and biases. Figure \ref{results} shows the comparison on the test set: we directly see an overall good agreement between the SDSS reference values and the values estimated with the RF algorithm, both for SFR and for $\mathrm{M}_\star$. This agreement shows that the RF algorithm is reasonably well trained. 

For the stellar mass values estimated from the RF algorithm (Fig. \ref{results} left panel), the scatter between the estimated and reference SDSS values is quantified through the variance: $\sigma^2_{\mathrm{M}_\star}=0.026$. The associated standard deviation of $\sigma_{\mathrm{M}_\star}=0.16$ dex which translates into an error of a factor $10^{\sigma_{\mathrm{M}_\star}}=1.45$ with respect to the reference value. For the SFR  (Fig. \ref{results} right panel), the scatter is larger and the variance is $\sigma^2_{\mathrm{SFR}}=0.145$. This gives a standard deviation of $\sigma_{\mathrm{SFR}}=0.38$ dex, and an error of a factor $10^{\sigma_{\mathrm{SFR}}}=2.40$. 

\subsection{Chasing the biases}

It is important to have precise results, with error bars estimated from the RF for both SFR and the $\mathrm{M}_\star$ values and it is of equal importance to have accurate, that is, unbiased, results.

We first investigate potential biases induced by the redshift dependence of the SFR and stellar mass in the redshift range, $0<z<0.3$,  of the training catalogue used in our study. We display in 
Fig.~\ref{along_red} the errors (defined as the difference between machine-learning estimated values and SDSS values), for $\mathrm{M}_\star$ (left panel) and SFR (right panel) for the galaxies of the test set. No obvious bias on redshift is observed. In the left panel of Fig.~\ref{along_red}  we notice a slight increase of the scatter for $\mathrm{M}_\star$ at very low redshifts. This is discussed in Sect.~\ref{trois}.

Another type of bias can be induced by the galaxy types. As we want a scatter of the same order for both passive and active galaxies, we compare the results of the RF algorithm as a function of the BPT classes provided in the SDSS MPA-JHU catalogue. To check that the BPT class is a reliable indicator of the galaxy type, we show in Fig.~\ref{probe_bpt} the main sequence diagram of galaxies (SFR vs. $\mathrm{M}_\star$ as provided in the SDSS catalogue) with their BPT classes from the SDSS catalogue. We see that the red contours of passive galaxies, BPT = -1, are well in the cloud of red and dead galaxies, the blue contours of star-forming galaxies, BPT = 1, are well along the main sequence, and the transitioning galaxies, BPT = 2, are well in the green valley. The positions in the BPT diagram can therefore be taken as a reliable indicator of galaxy type. In Fig.~\ref{BPT}, we show the results of the RF (same as those displayed in Fig.~\ref{results}), with the contour colours displaying the different BPT classes. The RF performs equally well for any type of galaxy and we do not observe any strong bias induced by galaxy type. Moreover, the scatter of the results depends only very slightly on galaxy type. For passive galaxies, BPT = -1, the scatter on $\mathrm{M}_\star$ tends to be reduced: we find $\sigma_{\mathrm{SFR}}=0.38$ dex and $\sigma_{\mathrm{M}_\star}=0.11$ dex. For active galaxies, BPT = 1, the inverse trend is seen and the scatter on the SFR tends to be reduced with a small increase of the scatter on $\mathrm{M}_\star$: we find $\sigma_{\mathrm{SFR}}=0.30$ dex and $\sigma_{\mathrm{M}_\star}=0.23$ dex. For transitioning galaxies, BPT = 2, we find roughly the same scatters as the overall ones on the global set given in Sect.~\ref{sec:test}, with $\sigma_{\mathrm{SFR}}=0.39$ dex and $\sigma_{\mathrm{M}_\star}=0.13$ dex. A summary of the different scatters is shown in Table~\ref{tableau}.

\begin{table*}
\centering
\begin{tabular}{c|c|c|c|c|c|c|}
\cline{2-7}
                       & \multicolumn{4}{c|}{ML with $z$} & ML without $z$ & Analytical \\ \cline{2-7} 
                       &  All   & Passive (BPT = -1)    &  Active (BPT = 1)   &  Green (BPT = 2)   & All & All \\ \hline
\multicolumn{1}{|c|}{$\sigma_{\mathrm{M}_\star}$} &  0.16  &    0.11 & 0.23    &  0.13   & 0.32 & 0.23 \\ \hline
\multicolumn{1}{|c|}{$\sigma_{\mathrm{SFR}}$} &  0.38   &  0.38   &  0.30   &  0.39   & 0.43 & 0.47 (active only) \\ \hline
\end{tabular}
 \caption{Summary of the different scatters obtained on the same test set with different methods.}
    \label{tableau}
\end{table*}

\subsection{Learning from the learning}

\begin{figure*}[!ht]
    \centering
    \includegraphics[width=0.33\textwidth]{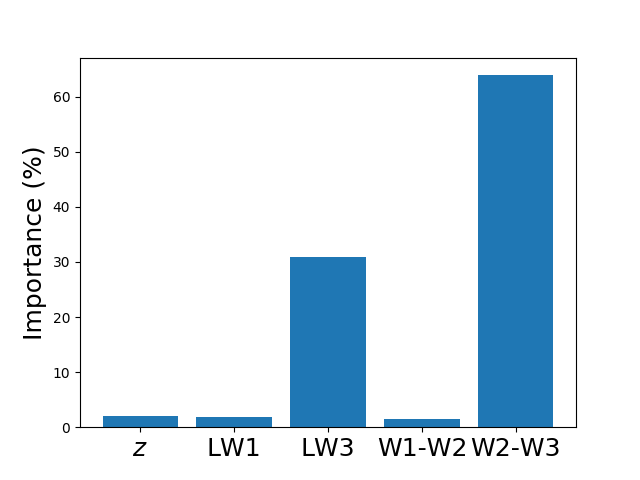}\includegraphics[width=0.33\textwidth]{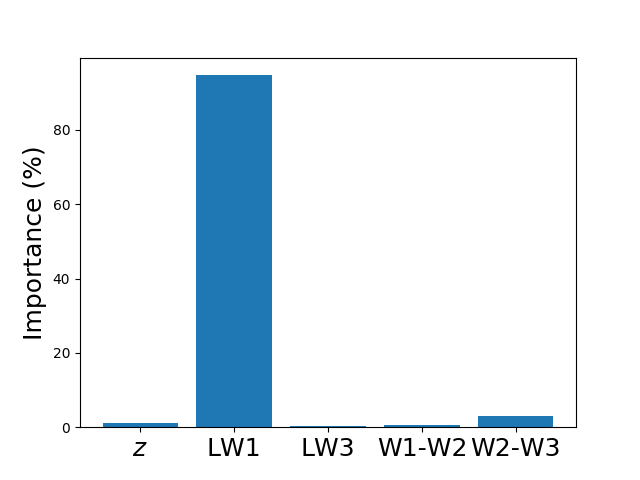}\includegraphics[width=0.33\textwidth]{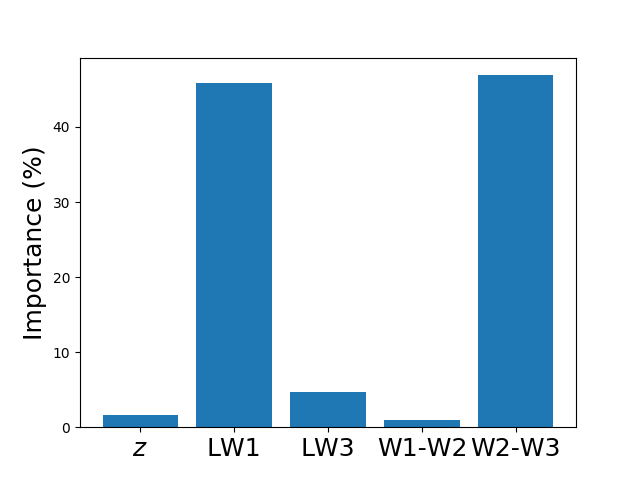}
    \caption{Feature importance during the RF training. Left: RF trained to perform $\mathrm{M}_\star$ estimates only. Middle: RF trained to perform SFR estimates only. Right: RF trained to estimate both SFR and $\mathrm{M}_\star$.}
    \label{importance}
\end{figure*}

\begin{figure}
    \centering
    \includegraphics[width=0.5\textwidth]{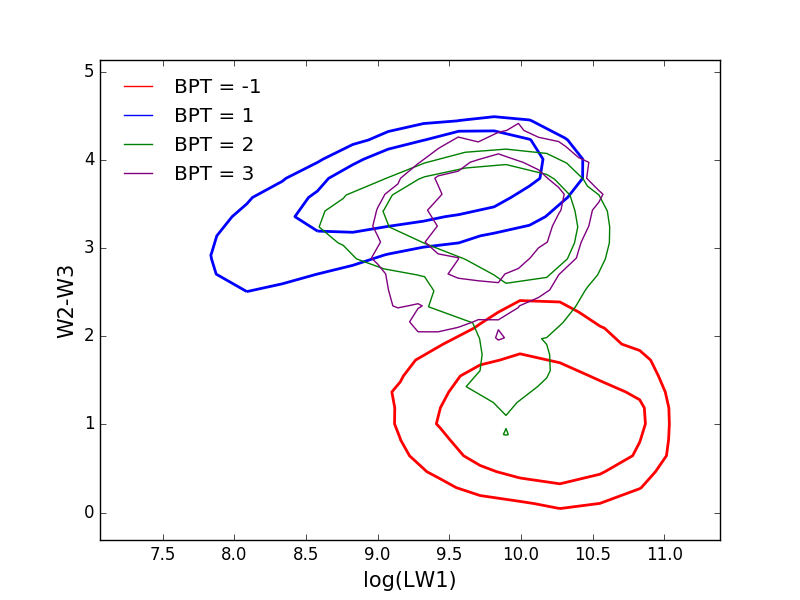}
    \caption{$1\sigma$ and $3\sigma$ iso-densities of the test set in the WISE colour-luminosity W2-W3/LW1 diagram. The line styles are the same as in Fig.~\ref{probe_bpt}.}
    \label{wise_collum}
\end{figure}

One advantage of the RF algorithm is that we can learn about the importance of the inputs during the training. An example of learning from the learning is to train the RF algorithm to estimate either SFR alone, $\mathrm{M}_\star$ alone, or both SFR and $\mathrm{M}_\star$ and to compare the importance of the features in each training (the left, middle, and right panel in Fig.~\ref{importance}, respectively). From the five input features, the first obvious tendency is the low impact of the redshift (only on the dependence on distance, as the redshift is also hidden in the luminosities LW1 and LW3) and of the colour W1-W2 on the training, whatever the outputs. For the estimation of $\mathrm{M}_\star$ alone, it is clear that the luminosity LW1, as expected, is the main feature used to train the RF, with a very slight contribution from the colour W2-W3. For the SFR estimation alone, two main features are used to train the RF, as expected: the luminosity LW3 and the colour W2-W3 used to segregate the two main populations of galaxies. The case where we train the RF to estimate both SFR and $\mathrm{M}_\star$ shows that the two main features are the luminosity LW1 and the colour W2-W3, with a slight contribution (of approx. 5\%) of the luminosity LW3. This indicates that the two quantities LW1 and W2-W3 are the most
efficient to classify and segregate galaxy populations (see also Fig.~\ref{wise_collum} where the two population of galaxies, i.e. red and blue, are very well separated.).

\begin{figure*}[!ht]
    \centering
    \includegraphics[width=0.5\textwidth]{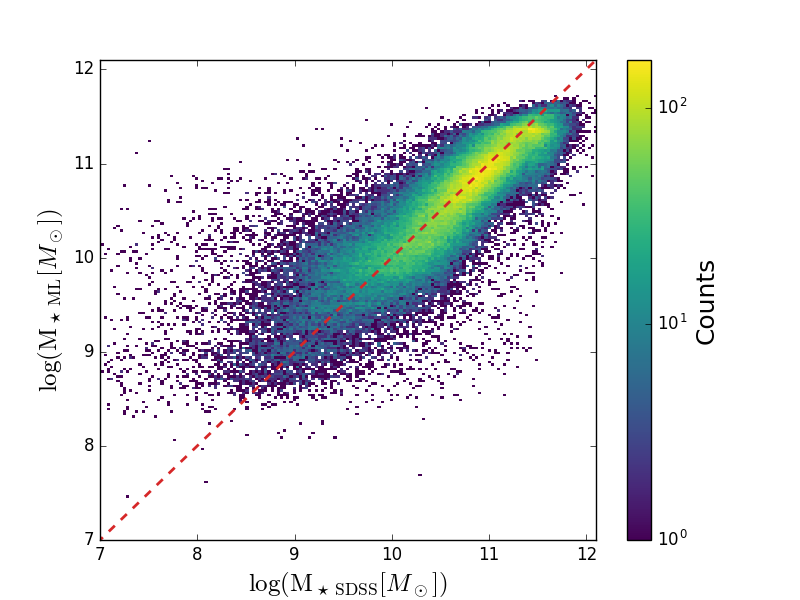}\includegraphics[width=0.5\textwidth]{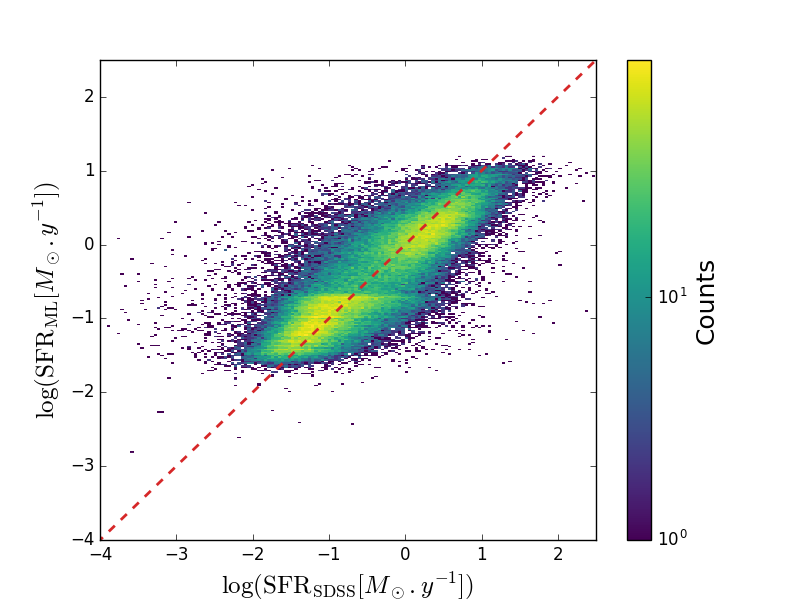}
    \caption{Results of the RF on the test sample (20\% of the entire sample), with only W1, W3, W1-W2 and W2-W3 in input and without information about the redshift. Left: $\mathrm{M}_\star$ estimate with the RF compared with $\mathrm{M}_\star$ from the SDSS MPA-JHU DR8 catalogue. Right: SFR estimate with the RF compared with SFR from the SDSS MPA-JHU DR8 catalogue.}
    \label{withoutz}
\end{figure*}

The estimated SFR and stellar mass from RF method rely on the redshift information. Redshifts are used to compute the luminosities and they impact the evolution of the mean global SFR over time. The need for redshifts to compute the $\mathrm{M}_\star$ and the SFR is very restrictive, as spectroscopic redshifts are hard to obtain and photometric ones are not always precise. We have tested the performance of the RF method without any redshift information. This implies that we do not perform any k-correction on the magnitudes and we do not compute the luminosities. The inputs are thus only the two magnitudes W1 and W3 and the two colours W1-W2 and W2-W3. In Fig.~\ref{withoutz}, we show the results on the test set. We find a scatter of $\sigma_{\mathrm{M}_\star}=0.32$ dex for the $\mathrm{M}_\star$ and a scatter of $\sigma_{\mathrm{SFR}}=0.43$ dex for the SFR estimation (compared to $\sigma_{\mathrm{M}_\star}=0.16$ dex and $\sigma_{\mathrm{SFR}}=0.38$ dex with the information about the redshift; see also Table~\ref{tableau}). The accuracy of the method is highly degraded and we then keep the redshift in our prior inputs.

\subsection{Applying to photometric redshift catalogues}

\begin{figure*}[!ht]
    \centering
    \includegraphics[width=0.5\textwidth]{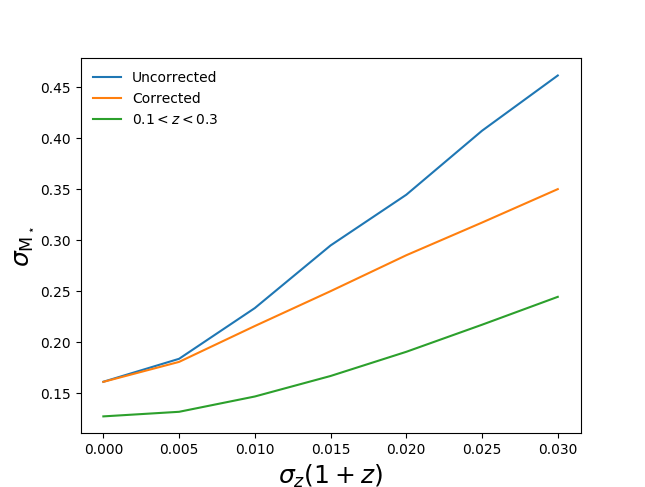}\includegraphics[width=0.5\textwidth]{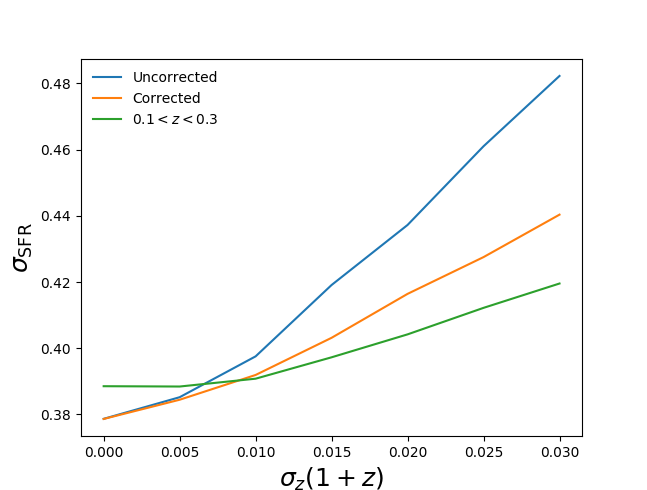}
    \caption{Evolution of the scatters of the estimated properties with the RF as a function of the redshift error. Left: Scatter for $\mathrm{M}_\star$ estimation. Right: Scatter for SFR estimation. The blue lines correspond to the scatters corresponding to the whole sample, regardless of the induced bias. The orange lines correspond to the scatters of bias-corrected properties following the laws in Fig.~\ref{bias_ind_ev}. The green lines correspond to the scatters in the redshift range $0.1<z<0.3$.}
    \label{scat_ev}
\end{figure*}

\begin{figure*}[!ht]
    \centering
    \includegraphics[width=0.5\textwidth]{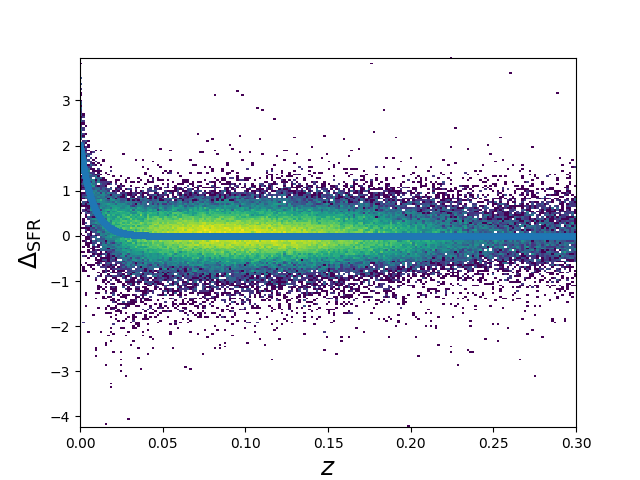}\includegraphics[width=0.5\textwidth]{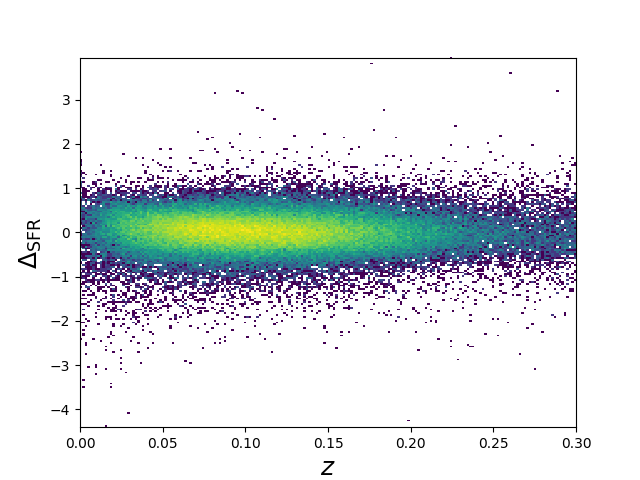}
    \caption{Example of bias induced by the redshift error. Left: For $\sigma_z(1+z)=0.015$ we show the errors on the SFR estimated values as a function of redshift. The blue line corresponds to the modeled bias. Right: Same errors as in the left panel but corrected for bias.}
    \label{bias_induced}
\end{figure*}

\begin{figure*}[!ht]
    \centering
    \includegraphics[width=0.5\textwidth]{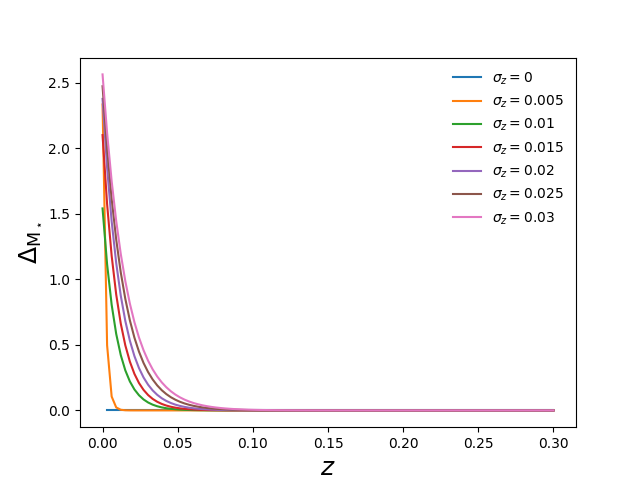}\includegraphics[width=0.5\textwidth]{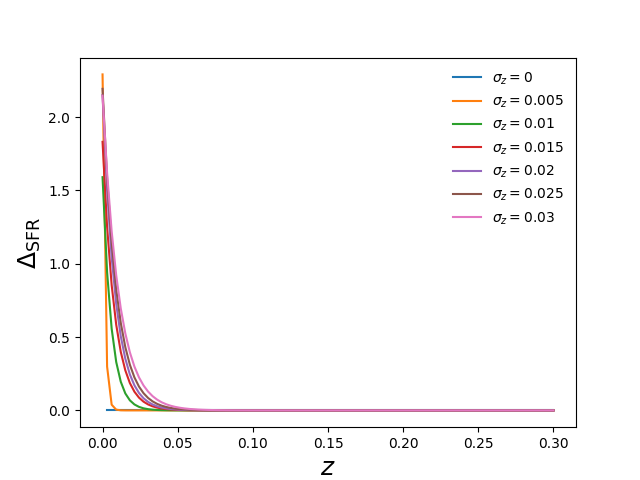}
    \caption{Evolution of the bias seen as a function of redshift, for different redshift errors (indicated by the colours). Left: Bias to correct for $\mathrm{M}_\star$ estimations. Right: Bias to correct for SFR estimations.}
    \label{bias_ind_ev}
\end{figure*}

The machine-learning algorithm is trained on a spectroscopic-redshift catalogue and can be applied to high-accuracy photometric-redshift catalogues. In order to test this, we add an error $\sigma_z(1+z)$, illustrative of errors from photometric redshifts, to the redshifts of the test sample. We then estimate the two properties, SFR and $\mathrm{M}_\star$. In Fig.~\ref{scat_ev}, we show in blue the evolution of the scatter of SFR and $\mathrm{M}_\star$ estimates as a function of $\sigma_z(1+z)$. We see a large increase of the scatter for both properties with decreasing redshift accuracy. This trend has two origins. On the one hand, the increase in redshift errors obviously impacts the SFR and $\mathrm{M}_\star$ estimates. On the other hand, an additional bias increases the errors on the estimates SFR and $\mathrm{M}_\star$. This is illustrated in the left panel of Fig.~\ref{bias_induced} showing the error on SFR for $\sigma_z(1+z)=0.015$. This bias can be corrected for. We have modelled it with a simple exponential $a\times\exp{-z/z_0}$ and show its evolution in Fig.~\ref{bias_ind_ev}. In Fig. \ref{scat_ev}, we show in orange the scatter on the bias-corrected properties. The scatters are significantly reduced, but still of $\sigma_{\mathrm{M}_\star}=0.35$ dex and $\sigma_{\mathrm{SFR}}=0.44$ dex at $\sigma_z(1+z) = 0.03$. If we focus on the SFR and $\mathrm{M}_\star$ estimates in the range $0.1<z<0.3$, the scatters (shown in green) are reduced down to more reasonable values such as $\sigma_{\mathrm{M}_\star}=0.24$ dex and $\sigma_{\mathrm{SFR}}=0.42$ dex at $\sigma_z(1+z) = 0.03$ (the accuracy expected for the photometric-redshift catalogue of Euclid). In this way, one could apply the present approach to the whole sky based on present or future photometric-redshift catalogues like WISExSCOS, DES, LSST, Pan-Starrs, or Euclid.

\section{Discussion}\label{discussion}

\subsection{Comparison}

Determination of SFR and  $\mathrm{M}_\star$  is an active topic, and other studies have provided analytical formulae, some of them also based on the WISE luminosities \citep[e.g.][]{wen2013, jarrett2013, cluver2014, cluver2017}. We compare the estimates of SFR and $\mathrm{M}_\star$ derived from the RF algorithm with those computed using different approaches but based on the same observables (WISE luminosities: LW1 and LW3). We focus on the $\mathrm{M}_\star$ estimates with the relation from \cite{wen2013}, using LW1:

\begin{equation}\label{coucou}
    \mathrm{log}\left({\mathrm{M}_\star}_\mathrm{Wen}\right) = 1.12\times\mathrm{log}\left(\mathrm{LW1}\right) - 0.04,
\end{equation}

and on the SFR estimates from \cite{cluver2014} using LW3, for star-forming galaxies:

\begin{equation}\label{coucou2}
    \mathrm{log}\left(\mathrm{SFR}_\mathrm{Cluver}\right) = 1.13\times\mathrm{log}\left(\mathrm{LW3}\right)-10.24.
\end{equation}

\begin{figure*}[!ht]
    \centering
    \includegraphics[width=0.5\textwidth]{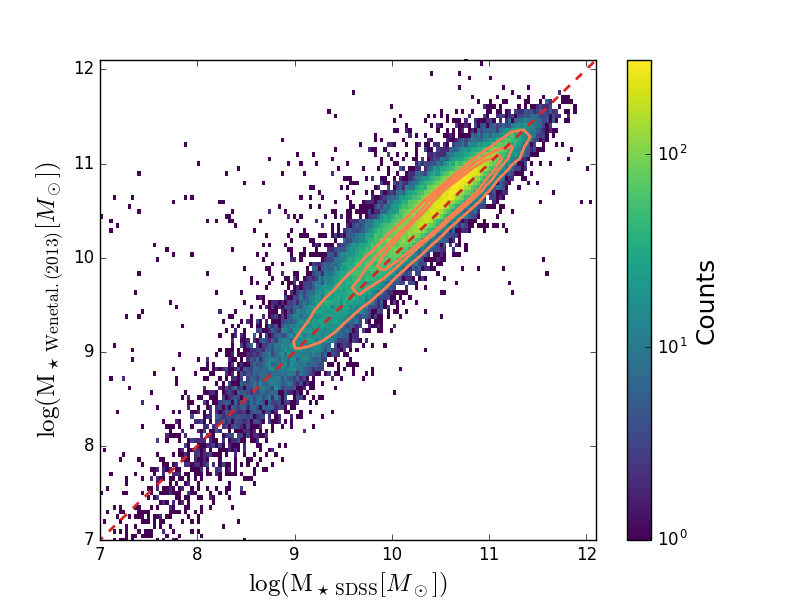}\includegraphics[width=0.5\textwidth]{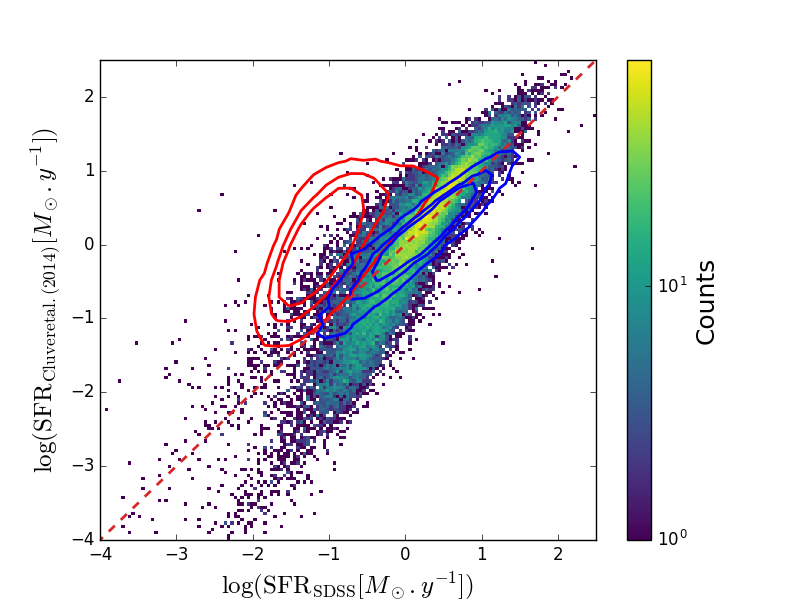}
    \caption{Comparisons on the test set. Left: $\mathrm{M}_\star$ computed with the method of \cite{wen2013}, using the luminosity in W1, compared with the SDSS masses. The contours show the 1, 3, and 5$\sigma$ isodensities of the RF results (Fig.~\ref{results}). Right: SFR computed only for star-forming galaxies with the method of \cite{cluver2014}, using the luminosity in W3, compared to SFR from the SDSS catalogue. The blue contours represent the results of the RF for the same population (blue contours in Fig.~\ref{BPT}), and the red contours represent the SFR estimation for passive galaxies computed with Cluver's formula. In both panels, the dashed line represents the one-to-one correlation.}
    \label{compare}
\end{figure*}

We compute $\mathrm{M}_\star$ with Eq.~\ref{coucou} for all galaxies and compare with the masses reported in the SDSS MPA-JHU catalogue (left panel of  Fig.~\ref{compare}). We also show the 1, 3, and 5$\sigma$ contours of the RF estimates for the same galaxies. This comparison shows the smaller scatter of the masses estimated with the RF algorithm, which is not surprising considering that we have five inputs compared to only one. For these specific sources, we find $\sigma_{{\mathrm{M}_\star}_\mathrm{Wen}}=0.23$ dex and $\sigma_{{\mathrm{M}_\star}_\mathrm{ML}}=0.16$ dex (see Table~\ref{tableau}).

The SFR values are computed only for star-forming galaxies (BPT = 1, to satisfy the conditions of \cite{cluver2014}) following Eq.~\ref{coucou2} and are compared with the SFR in the SDSS MPA-JHU catalogue (right panel of Fig.~\ref{compare}). The blue contours represent the results of the RF for the same population (blue contours in Fig.~\ref{BPT}), and the red contours represent the SFR estimates for passive galaxies (BPT = -1) computed with the \cite{cluver2014} formula (Eq.~\ref{coucou2}). We find a smaller scatter for the SFR estimations from RF; this is again expected since we use five inputs compared to one. We also show the limitation of the application domain of a linear relation between a luminosity and an SFR, in terms of its dependence on galaxy type (huge bias of the red contours). We find $\sigma_{{\mathrm{SFR}}_\mathrm{Cluver}}=0.47$ dex and $\sigma_{{\mathrm{SFR}}_\mathrm{ML}}=0.30$ dex  for active galaxies (see Table~\ref{tableau}), while for passive galaxies we have $\sigma_{{\mathrm{SFR}}_\mathrm{Cluver}}=0.49$ dex and $\sigma_{{\mathrm{SFR}}_\mathrm{ML}}=0.38$ dex and a bias (defined as the absolute difference of the means) of $b_\mathrm{Cluver}=0.93$ dex compared with $b_\mathrm{ML}=0.04$ dex.
 
\begin{figure*}
    \centering
    \includegraphics[width=0.33\textwidth]{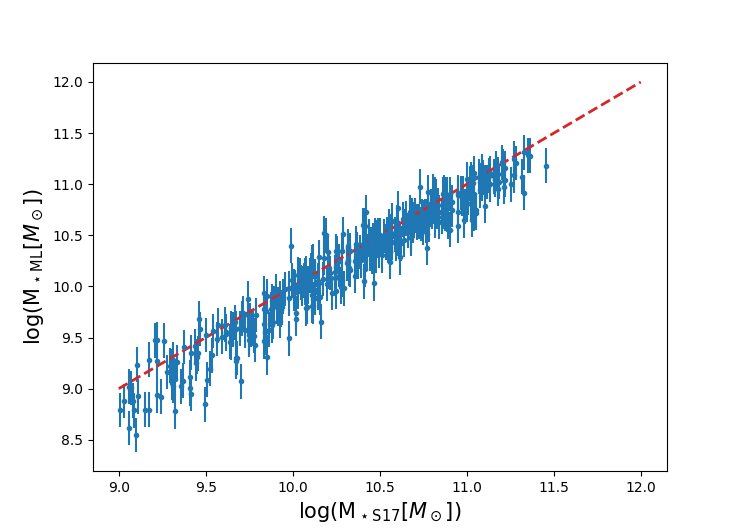}\includegraphics[width=0.33\textwidth]{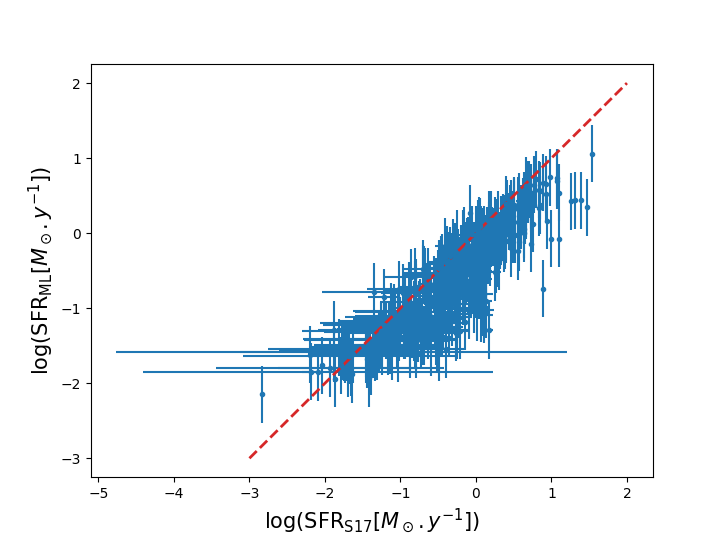}\includegraphics[width=0.33\textwidth]{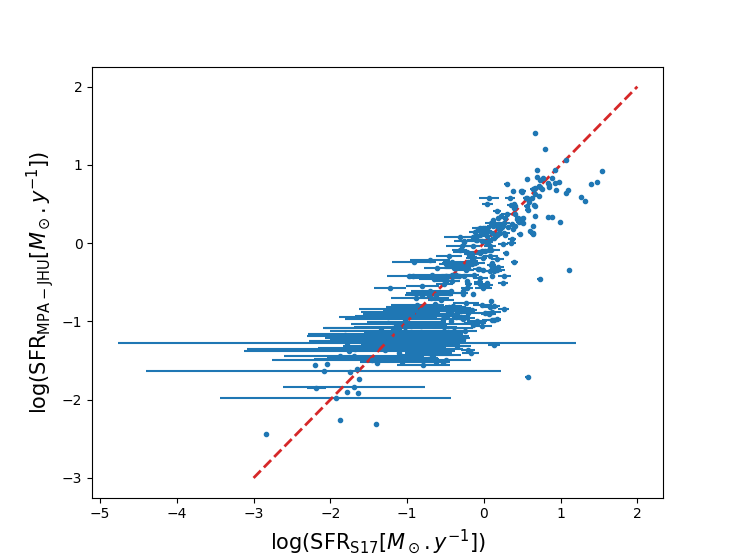}
    \caption{Comparison with the xCOLDGAS catalogue from \cite{saintonge17} for the sources that match only once with the AllWISE catalogue in a 6-arcsec radius. Left: $\mathrm{M}_\star$ computed with the RF compared with the $\mathrm{M}_\star$ provided by the catalogue (SDSS MPA-JHU values). Middle: SFR computed by the RF compared with the SFR provided by xCOLDGAS, computed with combined UV and IR data. Right: SFR given by the SDSS MPA-JHU DR8 catalogue compared with the SFR from xCOLDGAS, computed with combined UV and IR data.}
    \label{saintonge}
\end{figure*}

As a second comparison, we use a catalogue of galaxies with estimated SFRs calculated using an alternative method. An example is the extended version of the COLD GASS (CO Legacy Database for GASS) catalogue of nearby galaxies, xCOLD GASS\footnote{Publicly avaible at \url{http://www.star.ucl.ac.uk/xCOLDGASS/index.html}} \citep{saintonge17}. The sample contains 532 galaxies from SDSS selected in mass ($\mathrm{M}_\star> 10^9 \mathrm{M}_\odot$) that span a large range of SFR values and galaxy types, with IRAM-30m CO(1-0) observations, and as they say, \textit{``because the COLD GASS sample is large and unbiased, it serves as the perfect reference for studies of particular galaxy populations''}. \cite{saintonge17} provide ancillary information such as the SFR and $\mathrm{M}_\star$ which is computed with the method in \cite{janowiecki}, using a combination of UV from the Galaxy Evolution Explorer\footnote{\url{http://www.galex.caltech.edu}} (GALEX) and IR from WISE. We show in Fig.~\ref{saintonge} the results of the RF compared with the values of the xCOLDGASS catalogue (shown as S17), for the galaxies with only one association (no multiple source blending) with the AllWISE catalogue within a radius of 6 arcsec. Good  overall agreement is seen, and a small bias is seen in the SFR estimation, especially for passive galaxies. This bias is partially due to the way SFR is estimated. Here, we take the SDSS MPA-JHU catalogue estimation as reference to train the model, and we compare to their SFR computed with both IR and UV data. The reliability of the values estimated with the RF is directly correlated with the reliability of the values chosen as outputs, that is, the SDSS MPA-JHU values. We can also see the bias between the SDSS MPA-JHU values and the UV+IR estimation in the panel right of Fig.~\ref{saintonge}, and show that maybe unobscured UV SFR is not seen in WISE or in SDSS. The method could be applied to any other value-added catalogue of galaxies chosen as outputs of the machine-learning training, if more robust SFR or $\mathrm{M}_\star$ estimators (via multi-wavelength 
proxies, i.e. IR+UV) were to teach more accurate predictions (see Fig.~\ref{saintonge}).

\subsection{Limitation}\label{trois}
 
The results of the RF algorithm agree well with previous works, in particular those of \cite{wen2013} and \cite{cluver2014}. The domain of application of all three studies in terms of galaxies and more precisely in terms of redshifts is similar (mean redshift at around $z=0.15$). Comparison with other available catalogues providing SFR and $\mathrm{M}_\star$ cannot be possible when the sources under consideration are too different from the domain of application of the RF algorithm, that is, its learned model. We focus on two extreme cases of domains of application where the machine learning cannot provide reasonable estimates. On the one hand, we consider nearby galaxies with redshifts $z<0.01$ and on the other hand high-redshift galaxies, up to $z=8$.

A sample of nearby ($z<0.01$) star-forming galaxies was constructed combining a Spitzer catalogue SINGS \citep{kennicutt2003} and a Herschel Space Observatory catalogue KINGFISH \citep{kennicutt2011}. This SINGS/KINGFISH catalogue of 79 sources was used by \cite{cluver2017} to successfully fit the relation between SFR and LW3. In this catalogue, the redshift domain ($z<0.01$) implies that the galaxies are resolved and therefore the use of the WISE Atlas Images is needed to accurately measure the fluxes of the objects \citep{saintonge17, cluver2017}. The effect of a miscomputation of the IR fluxes for very-low-redshift resolved galaxies can be seen in Fig.~\ref{along_red} as a higher scatter in $\mathrm{M}_\star$ estimations for very-low-redshift galaxies ($z<0.01$).

An example of distant galaxies is the COSMOS2015 catalogue\footnote{\url{http://cosmos.astro.caltech.edu}} \citep{laigle}, which provides apparent magnitudes in 30 bands for approximately half a million objects up to redshifts $z=8$. It also provides photometric redshifts, SFRs, and stellar masses all computed using the code \texttt{LEPHARE}\footnote{\url{http://www.cfht.hawaii.edu/~arnouts/LEPHARE/lephare.html}}. Here again we cannot directly apply our RF algorithm, that is, its learned model, and compare our estimates with those of the COSMOS catalogue. The main issue here is the resolution of WISE of 6 arcsec; too many COSMOS sources are associated with a WISE galaxy inside the WISE beam and therefore correct association with the AllWISE catalogue is impossible. 

Finally, as the RF algorithm has been trained on galaxies up to $z=0.3$, using it beyond this redshift limit will lead to strong biases as the training did not include such sources. We conclude that the application domain of the method is $0.01<z<0.3$.

\subsection{Application}

As shown in statistical studies of galaxy populations in large galaxy surveys (e.g. SDSS \citep{sdss}, VIMOS Public Extragalactic Redshift Survey (VIPERS) \citep{vipers}), most galaxies residing in dense environments of the cosmic web (galaxy clusters or inner parts of cosmic filaments) are passive and red and dead \citep{malavasi1, malavasi2, kraljic}. The information about the activity of galaxies is indeed used to detect galaxy clusters using their red sequence, originally exhibited by \cite{gladders} (e.g. \cite{rykoff} who developed the RedMapper algorithm).

Similarly to estimating the activity of galaxies by computing specific SFR (which illustrates the efficiency of a galaxy in forming stars), the distance to the main sequence on an SFR vs. $\mathrm{M}_\star$ plot (translated into the colours on the Fig.~\ref{d2ms}) informs us about the activity of a galaxy. Since this distance, called d2ms, is directly related to how far galaxies are from being star-forming, we prefer to introduce the term passivity rather than activity. The more red the point, the more distant from the main sequence of star-forming galaxies and the more passive the galaxies are. This quantity (quite efficiently estimated by our method) can be a very useful property to segregate populations of star-forming, green, or passive galaxies. We already successfully used this estimate in \cite{bonjean} to characterise the properties of galaxies in a cosmic bridge between two galaxy clusters.

\begin{figure*}[!ht]
    \centering
    \includegraphics[width=0.33\textwidth]{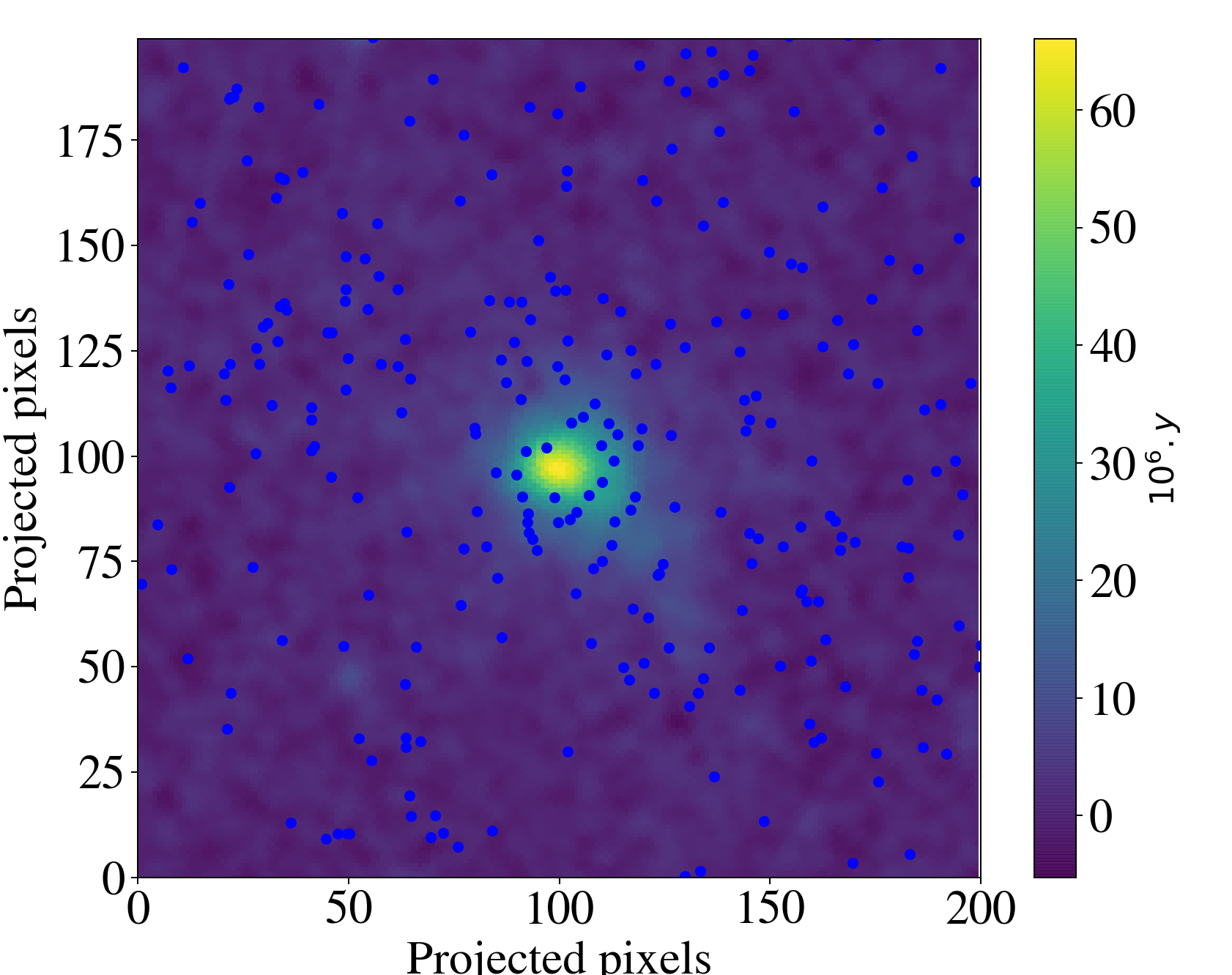}\includegraphics[width=0.33\textwidth]{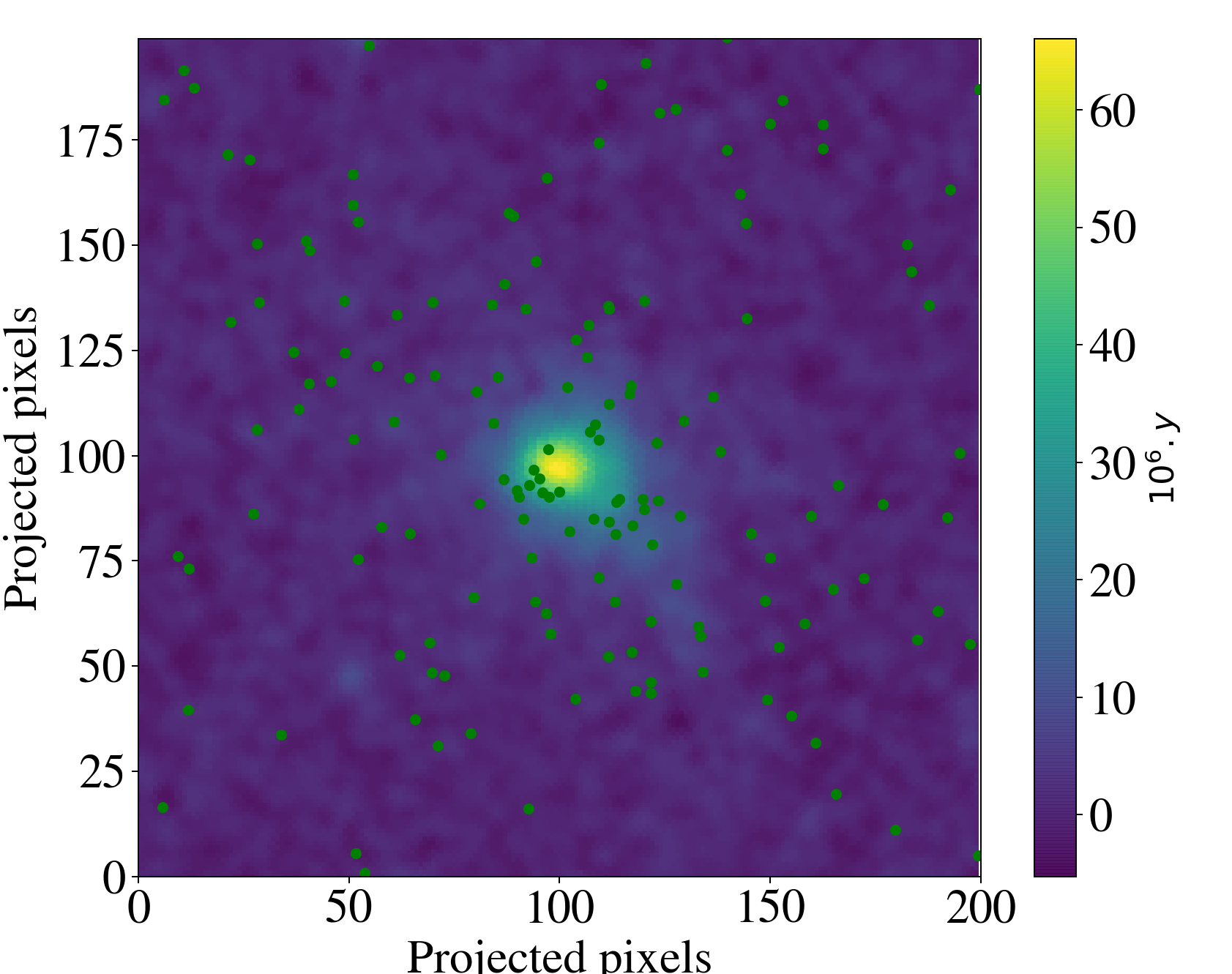}\includegraphics[width=0.33\textwidth]{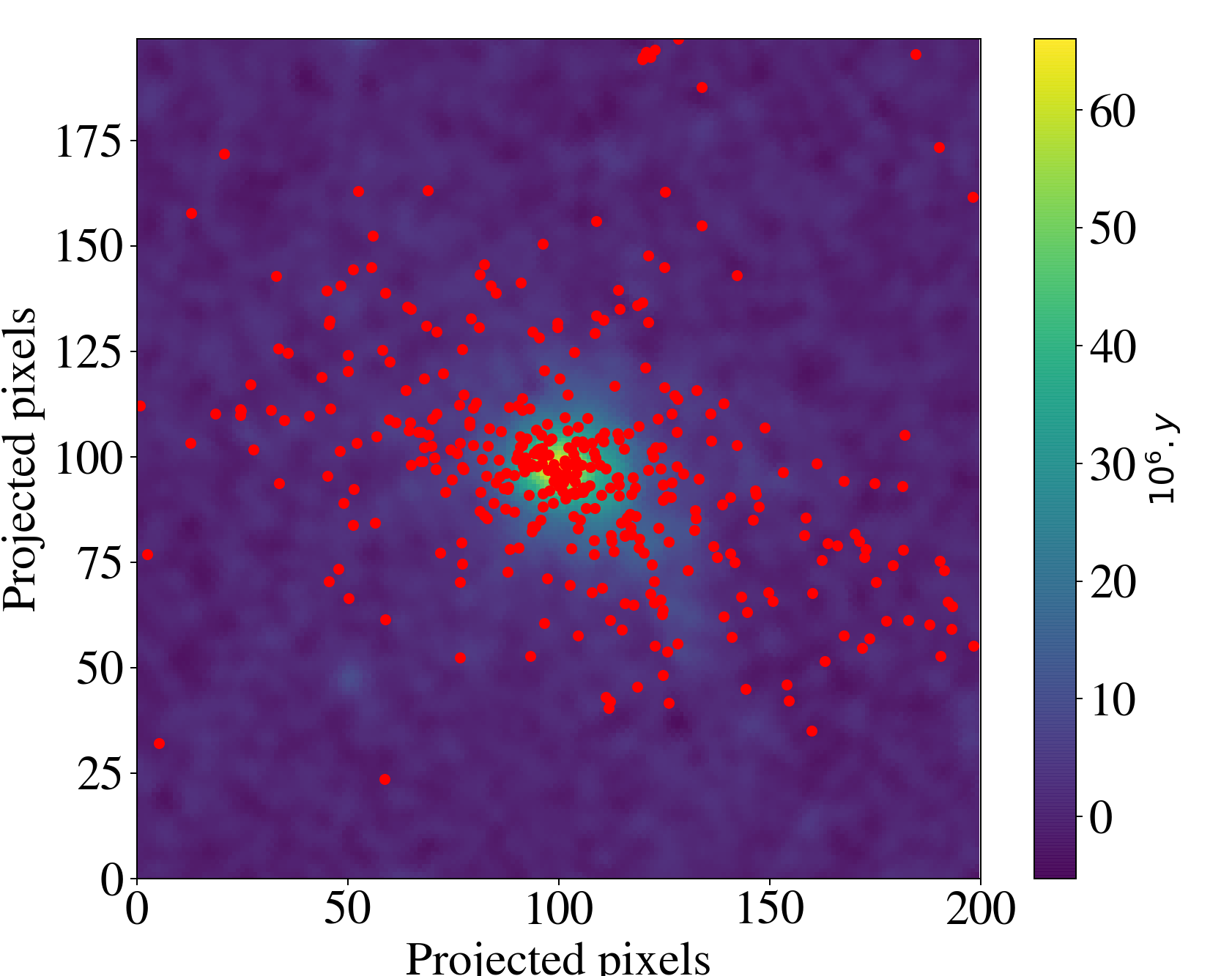}
    \caption{Example of application of our method for galaxy populations in a $10^{\circ}\times10^{\circ}$ field centred on the Coma cluster at $z=0.0231$. We have estimated SFR and $\mathrm{M}_\star$ on 2MPZ union WISExSCOS galaxies with $0.01<z<0.05$. Based on the ``passivity'' cut (d2ms parameter) to segregate population, we select blue, green, and red galaxies corresponding to d2ms<0.4, 0.4<d2ms<1.25, and d2ms>1.25, respectively. The galaxies are overlaid on the thermal Sunyaev-Zel'dovich MILCA map from \textit{Planck} \citep{szmap}.}
    \label{coma}
\end{figure*}

We show in Fig.~\ref{coma} another simple illustration with the case study of galaxy population in the field of the Coma cluster at $z=0.0231$. As in \cite{bonjean}, we used the union between the 2MPZ and the WISExSCOS catalogues of photometric redshifts as a basis of galaxy catalogues. Extracting a region of 15 degrees around the Coma cluster, we select galaxies in the range $0.01<z<0.05$, and separate their population by performing cuts in the distance to the main sequence d2ms (see Fig.~\ref{d2ms}), computed from the SFR and $\mathrm{M}_\star$. The cuts in the d2ms correspond to d2ms<0.4, 0.4<d2ms<1.25, and d2ms>1.25 for blue, green, and blue galaxies, respectively (defined using Fig.~\ref{d2ms}). We clearly see the overdensity of red galaxies, as expected, clustered in the centre of the Coma cluster, and a more uniform distribution of blue galaxies around the Coma cluster and in the field. The galaxies are overlaid on the thermal Sunyaev-Zel'dovich MILCA map from {\ Planck} \citep{szmap}. This illustrates well the complexity of the galaxy dynamics inside the Coma cluster.

\section{Summary}\label{summary}

Determining star-formation activity proxies such as the SFR and  $\mathrm{M}_\star$  from UV, optical, or IR luminosities relies on complex modelling and on priors on galaxy properties, and does not accurately describe the passive galaxies, which are of particular interest in studying large-scale structures. We have developed a method based on machine learning to estimate the SFR and $\mathrm{M}_\star$ of galaxies, in the redshift range $0.01 < z < 0.3$, over the whole usable sky when their redshifts are known. The algorithm is trained on the redshift $z$, the luminosities LW1 and LW3, and the colours W1-W2 and W2-W3. These input properties permit a very efficient segregation between the different galaxy types and computation of the stellar masses and SFR. As outputs to train the algorithm, we have chosen the SDSS MPA-JHU DR8 values of SFR and $\mathrm{M}_\star$ as reference values. However, the method can be adapted to any catalogue of value-added SFR and $\mathrm{M}_\star$ cross-matched with the AllWISE catalogue.
With our method, we obtain typical errors of $\sigma_{\mathrm{SFR}}=0.38$ dex and $\sigma_{\mathrm{M}_\star}=0.16$ dex, independently of galaxy type, and unbiased with respect to redshift in the range $z<0.3$. 

In future works, we will extend the redshift range of the application up to $z\sim0.5$, and possibly beyond. In doing so, our method will be useful to characterise cosmic structures, and can therefore be applied, for example, to studies of the dependence of galaxy populations on environments or galaxy cluster/cosmic filament detection in the context of future large galaxy surveys.




\begin{acknowledgements}
The authors thank the anonymous referee for her/his useful comments. The authors acknowledge fruitful discussions with M. Bilicki, A. Decelle, M. Huertas-Company, and H.J. McCraken. We especially thank M. Cluver for providing her data on the SINGS/KINGFISH sample.
This publication used data products from the Wide-field Infrared Survey Explorer, which is a joint project of the University of California, Los Angeles, and the Jet Propulsion Laboratory/California Institute of Technology, funded by NASA. This research made use of Astropy, the community-developed core Phyton package \citep{astropy}. This project has received funding from the European Research Council (ERC) under the European Union's Horizon 2020 research and innovation programme grant agreement ERC-2015-AdG 695561.
\end{acknowledgements}


\begin{thebibliography}{}
        \bibitem[Aghanim et al.(2015)]{aghanim2015} Aghanim, N., Hurier, G., Diego, J.-M., et al.\ 2015, \aap, 580, A138
        \bibitem[Alatalo et al.(2014)]{alatalo2014} Alatalo, K., Cales, S.~L., Appleton, P.~N., et al.\ 2014, \apjl, 794, L13 
          \bibitem[Astropy Collaboration et al.(2013)]{astropy} Astropy Collaboration, Robitaille, T.~P., Tollerud, E.~J., et al.\ 2013, \aap, 558, A33 
            \bibitem[Baldwin et al.(1981)]{bpt} Baldwin, J.~A., Phillips, M.~M., \& Terlevich, R.\ 1981, \pasp, 93, 5
              \bibitem[Balogh et al.(1999)]{balogh1999} Balogh, M.~L., Morris, S.~L., Yee, H.~K.~C., Carlberg, R.~G., \& Ellingson, E.\ 1999, \apj, 527, 54
                \bibitem[Bilicki et al.(2014)]{bilicki2014} Bilicki, M., Jarrett, T.~H., Peacock, J.~A., Cluver, M.~E., \& Steward, L.\ 2014, \apjs, 210, 9
                  \bibitem[Bilicki et al.(2016)]{bilicki2016} Bilicki, M., Peacock, J.~A., Jarrett, T.~H., et al.\ 2016, \apjs, 225, 5 
                    \bibitem[Brinchmann et al.(2004)]{brinchmann} Brinchmann, J., Charlot, S., White, S.~D.~M., et al.\ 2004, \mnras, 351, 1151
                      \bibitem[Bonjean et al.(2018)]{bonjean} Bonjean, V., Aghanim, N., Salom{\'e}, P., Douspis, M., \& Beelen, A.\ 2018, \aap, 609, A49 
                        \bibitem[Bruzual A.(1983)]{bruzual1983} Bruzual A., G.\ 1983, \apj, 273, 105 
        \bibitem[Bruzual \& Charlot(2003)]{bruzual} Bruzual, G., \& Charlot, S.\ 2003, \mnras, 344, 1000 
          \bibitem[Calzetti et al.(1994)]{calzetti1994} Calzetti, D., Kinney, A.~L., \& Storchi-Bergmann, T.\ 1994, \apj, 429, 582 
          \bibitem[Calzetti et al.(2007)]{calzetti} Calzetti, D., Kennicutt, R.~C., Engelbracht, C.~W., et al.\ 2007, \apj, 666, 870 \bibitem[Chabrier(2003)]{chabrier} Chabrier, G.\ 2003, \pasp, 115, 763 
            \bibitem[Cluver et al.(2014)]{cluver2014} Cluver, M.~E., Jarrett, T.~H., Hopkins, A.~M., et al.\ 2014, \apj, 782, 90
              \bibitem[Cluver et al.(2017)]{cluver2017} Cluver, M.~E., Jarrett, T.~H., Dale, D.~A., et al.\ 2017, \apj, 850, 68 
                \bibitem[Cutri et al.(2013)]{cutri2013} Cutri, R.~M., Wright, E.~L., Conrow, T., et al.\ 2013, Explanatory Supplement to the AllWISE Data Release Products, by R.~M.~Cutri et al.~,  
                \bibitem[Dubois et al.(2013)]{dubois2013} Dubois, Y., Gavazzi, R., Peirani, S., \& Silk, J.\ 2013, \mnras, 433, 3297\bibitem[Dom{\'{\i}}nguez S{\'a}nchez et al.(2018)]{domingez} Dom{\'{\i}}nguez S{\'a}nchez, H., Huertas-Company, M., Bernardi, M., Tuccillo, D., \& Fischer, J.~L.\ 2018, \mnras, 476, 3661 
                  \bibitem[Elbaz et al.(2007)]{elbaz} Elbaz, D., Daddi, E., Le Borgne, D., et al.\ 2007, \aap, 468, 33
                    \bibitem[Gladders \& Yee(2000)]{gladders} Gladders, M.~D., \& Yee, H.~K.~C.\ 2000, \aj, 120, 2148 
                      \bibitem[Haas \& Anders(2010)]{haas} Haas, M.~R., \& Anders, P.\ 2010, \aap, 512, A79
                        \bibitem[Ho, Tin Kam(1995)]{ho1998}
                          Ho, Tin Kam\ 1995, Random Decision Forests. Proceedings of the 3rd International Conference on Document Analysis and Recognition, Montreal, QC, 14–16 August 1995. pp. 278–282.
                        \bibitem[Huertas-Company et al.(2015)]{marc2015} Huertas-Company, M., Gravet, R., Cabrera-Vives, G., et al.\ 2015, \apjs, 221, 8\bibitem[Janowiecki et al.(2017)]{janowiecki} Janowiecki, S., Catinella, B., Cortese, L., et al.\ 2017, \mnras, 466, 4795 
                          \bibitem[Jarrett et al.(2013)]{jarrett2013} Jarrett, T.~H., Masci, F., Tsai, C.~W., et al.\ 2013, \aj, 145, 6 
                            \bibitem[Kauffmann et al.(2003)]{kauffmann} Kauffmann, G., Heckman, T.~M., White, S.~D.~M., et al.\ 2003, \mnras, 341, 33
                              \bibitem[Kennicutt(1998)]{kennicutt1998} Kennicutt, R.~C., Jr.\ 1998, \araa, 36, 189 
                                \bibitem[Kennicutt et al.(2003)]{kennicutt2003} Kennicutt, R.~C., Jr., Armus, L., Bendo, G., et al.\ 2003, \pasp, 115, 928 
                                  \bibitem[Kennicutt et al.(2009)]{kennicutt2009} Kennicutt, R.~C., Jr., Hao, C.-N., Calzetti, D., et al.\ 2009, \apj, 703, 1672-1695
                                    \bibitem[Kennicutt et al.(2011)]{kennicutt2011} Kennicutt, R.~C., Calzetti, D., Aniano, G., et al.\ 2011, \pasp, 123, 1347 
                                      \bibitem[Kennicutt \& Evans(2012)]{kennicutt2012} Kennicutt, R.~C., \& Evans, N.~J.\ 2012, \araa, 50, 531
                                        \bibitem[Krakowski et al.(2016)]{krakowski} Krakowski, T., Ma{\l}ek, K., Bilicki, M., et al.\ 2016, \aap, 596, A39
                                          \bibitem[Kraljic et al.(2018)]{kraljic} Kraljic, K., Arnouts, S., Pichon, C., et al.\ 2018, \mnras, 474, 547 
                                            \bibitem[Kroupa(2001)]{kroupa} Kroupa, P.\ 2001, \mnras, 322, 231 
                                              \bibitem[Lagache et al.(2005)]{lagache2005} Lagache, G., Puget, J.-L., \& Dole, H.\ 2005, \araa, 43, 727 
                                                \bibitem[Laigle et al.(2016)]{laigle} Laigle, C., McCracken, H.~J., Ilbert, O., et al.\ 2016, \apjs, 224, 24 
                                                  \bibitem[Leger \& Puget(1984)]{pah} Leger, A., \& Puget, J.~L.\ 1984, \aap, 137, L5 
                                                    \bibitem[Lucie-Smith et al.(2018)]{rfsimu} Lucie-Smith, L., Peiris, H.~V., Pontzen, A., \& Lochner, M.\ 2018, arXiv:1802.04271
                                                      \bibitem[Malavasi et al.(2017a)]{malavasi1} Malavasi, N., Arnouts, S., Vibert, D., et al.\ 2017, \mnras, 465, 3817 
                                                        \bibitem[Malavasi et al.(2017b)]{malavasi2} Malavasi, N., Pozzetti, L., Cucciati, O., et al.\ 2017, \mnras, 470, 1274 
                                                          \bibitem[Moore et al.(1996)]{moore1995} Moore, B., Katz, N., Lake, G., Dressler, A., \& Oemler, A.\ 1996, \nat, 379, 613
                                                            \bibitem[Pashchenko et al.(2018)]{pashchenko} Pashchenko, I.~N., Sokolovsky, K.~V., \& Gavras, P.\ 2018, \mnras, 475, 2326 
                                                              \bibitem[Pedregosa et al.(2011)]{sklearn} Pedregosa, F., Varoquaux, G., et al.\ 2011, Journal of Machine Learning Research, 12, pp. 2825-2830
                                                                \bibitem[Peng et al.(2015)]{peng2015} Peng, Y., Maiolino, R., \& Cochrane, R.\ 2015, \nat, 521, 192
                                                                  \bibitem[Planck Collaboration et al.(2016a)]{planckcosmo} Planck Collaboration, Ade, P.~A.~R., Aghanim, N., et al.\ 2016a, \aap, 594, A13 
                                                                    \bibitem[Planck Collaboration et al.(2016b)]{szmap} Planck Collaboration, Aghanim, N., Arnaud, M., et al.\ 2016b, \aap, 594, A22
                                                                      \bibitem[Rykoff et al.(2014)]{rykoff} Rykoff, E.~S., Rozo, E., Busha, M.~T., et al.\ 2014, \apj, 785, 104 
                                                                        \bibitem[Saintonge et al.(2011)]{saintonge11} Saintonge, A., Kauffmann, G., Kramer, C., et al.\ 2011, \mnras, 415, 32 
                                                                          \bibitem[Saintonge et al.(2017)]{saintonge17} Saintonge, A., Catinella, B., Tacconi, L.~J., et al.\ 2017, \apjs, 233, 22
                                                                            \bibitem[Salim et al.(2007)]{salim2007} Salim, S., Rich, R.~M., Charlot, S., et al.\ 2007, \apjs, 173, 267
                                                                              \bibitem[Salpeter(1955)]{salpeter} Salpeter, E.~E.\ 1955, \apj, 121, 161
                                                                                
                                                                                \bibitem[Scodeggio et al.(2018)]{vipers} Scodeggio, M., Guzzo, L., Garilli, B., et al.\ 2018, \aap, 609, A84
                                                                                  \bibitem[Siudek et al.(2018a)]{vipers2018a} Siudek, M., Ma{\l}ek, K., Pollo, A., et al.\ 2018a, arXiv:1805.09904 
        \bibitem[Siudek et al.(2018b)]{vipers2018b} Siudek, M., Ma{\l}ek, K., Pollo, A., et al.\ 2018b, arXiv:1805.09905 
          \bibitem[Tuccillo et al.(2018)]{tuccillo} Tuccillo, D., Huertas-Company, M., Decenci{\`e}re, E., et al.\ 2018, \mnras, 475, 894 
            \bibitem[Ucci et al.(2018)]{ucci} Ucci, G., Ferrara, A., Pallottini, A., \& Gallerani, S.\ 2018, \mnras, 
              \bibitem[delli Veneri et al.(2018)]{veneri2018} delli Veneri, M., Cavuoti, S., Brescia, M., Riccio, G., \& Longo, G.\ 2018, arXiv:1805.06338 
                \bibitem[Viquar et al.(2018)]{viqar2018} Viquar, M., Basak, S., Dasgupta, A., Agrawal, S., \& Saha, S.\ 2018, arXiv:1804.05051 
                  \bibitem[Wen et al.(2013)]{wen2013} Wen, X.-Q., Wu, H., Zhu, Y.-N., et al.\ 2013, \mnras, 433, 2946 
                    \bibitem[Werner et al.(2004)]{spitzer} Werner, M.~W., Roellig, T.~L., Low, F.~J., et al.\ 2004, \apjs, 154, 1 
                      \bibitem[Wright et al.(2010)]{wise} Wright, E.~L., Eisenhardt, P.~R.~M., Mainzer, A.~K., et al.\ 2010, \aj, 140, 1868-1881
                        \bibitem[York et al.(2000)]{sdss} York, D.~G., Adelman, J., Anderson, J.~E., Jr., et al.\ 2000, \aj, 120, 1579 
                          
\end{thebibliography}
\end{document}